\documentclass[epj,nopacs]{svjour}
\pdfoutput=1
\usepackage[utf8]{inputenc}
\usepackage{color,graphicx,hyperref,cite,amssymb,amsmath,slashed}
\usepackage[normalem]{ulem}
\hypersetup{colorlinks=true,
  linkcolor=blue,citecolor=magenta,filecolor=magenta,urlcolor=cyan}
\newcommand{\be}{\begin{equation}}
\newcommand{\ee}{\end{equation}}
\def\bsp#1\esp{\begin{split}#1\end{split}}
\def\etc{{\it etc.}}
\def\ie{{\it i.e.}}

\newcommand{\madanalysis}{{\sc MadAnalysis~5}}
\newcommand{\madgraph}{{\sc MG5\_aMC}}
\newcommand{\madspin}{{\sc MadSpin}}
\newcommand{\madwidth}{{\sc MadWidth}}
\newcommand{\pythia}{{\sc Pythia}}
\newcommand{\delphes}{{\sc Delphes~3}}
\newcommand{\fr}{{\sc FeynRules}}
\newcommand{\fastjet}{{\sc FastJet}}
\newcommand{\lhapdf}{{\sc LHAPDF}}
\newcommand{\rooot}{{\sc Root}}

\newcommand{\gpp}{{\sc G++}}

\begin{document}

\title{
  Reinterpreting the results of the LHC with \madanalysis: uncertainties and
  higher-luminosity estimates
}

\author{
  Jack Y. Araz\inst{1}\thanks{\color{blue}jack.araz@concordia.ca},
  Mariana Frank\inst{1}\thanks{\color{blue}mariana.frank@concordia.ca} and
  Benjamin Fuks\inst{2,3}\thanks{\color{blue}fuks@lpthe.jussieu.fr}
}

\institute{
    Concordia University 7141 Sherbrooke St. West, Montr\'{e}al, QC,
    Canada H4B 1R6\label{addr1}
  \and 
    Laboratoire de Physique Th\'eorique et Hautes Energies (LPTHE),
    UMR 7589, Sorbonne Universit\'e et CNRS, 4 place Jussieu,
    75252 Paris Cedex 05, France\label{addr2}
  \and
    Institut Universitaire de France, 103 boulevard Saint-Michel, 75005 Paris,
    France\label{addr3}
}

\date{}

\abstract{
  The \madanalysis\ framework can be used to assess the potential of various LHC
  analyses for unraveling any specific new physics signal. We present an extension of the
  LHC reinterpretation capabilities of the programme allowing for the inclusion
  of theoretical and systematical uncertainties on the signal in the
  reinterpretation procedure. We have implemented extra methods dedicated to the
  extrapolation of the impact of a given analysis to higher luminosities,
  including various options for the treatment of the errors. As an application,
  we study three classes of new physics models. We first focus on a simplified
  model in which the Standard Model is supplemented by a gluino and a
  neutralino. We show that uncertainties could in particular degrade the bounds
  by several hundreds of GeV when considering 3000~fb$^{-1}$ of future LHC data.
  We next investigate another supersymmetry-inspired simplified model, in which
  the Standard Model is extended by a first generation squark species and a
  neutralino. We reach similar conclusions. Finally, we study a class of
  $s$-channel dark matter setups and compare the expectation for two types of
  scenarios differing in the details of the implementation of the mediation
  between the dark and Standard Model sectors.}

\titlerunning{Uncertainties and high-luminosity estimates with \madanalysis~recasting}
\authorrunning{J.Y.~Araz {\it et al.}}

\maketitle
\flushbottom

\section{Introduction}\label{sec:intro}
The discovery of the Higgs boson has accomplished one of the long awaited
objectives of the LHC physics programme and confirmed our
understanding of the fundamental laws of nature. However, the concrete
realisation of the electroweak symmetry breaking mechanism remains unexplained
and no evidence for physics beyond the Standard Model (SM), whose existence is
motivated by the SM theoretical inconsistencies and limitations, has emerged
from data. There are two classes of possible explanations as to why the
associated new particles and/or interactions have escaped detection so far. The
first one is that the new states are too heavy and/or the new interactions too
feeble to be observed with present collider reaches. Alternatively, new
particles may be hiding just around the corner, but lie in a specific
configuration (like being organised in a
compressed spectrum) that renders their discovery challenging. The possible
observation of any new phenomena therefore is the foremost goal of the
future LHC runs, including in particular the LHC Run~3, to be started in two
years, and the high-luminosity operations planned to begin in half a decade.

In order to investigate whether new physics could be present in existing data,
several groups have developed and maintained public software dedicated to the
reinterpretation of the results at the LHC~\cite{Kraml:2013mwa,Drees:2013wra,%
Dumont:2014tja,Buckley:2010ar,Balazs:2017moi}. In practice, these tools rely
on predictions detailing how the different signal regions of given LHC analyses
are populated to derive the potential of these searches for its observation.
However, signal uncertainties are in general ignored by users in
this procedure, although they could sometimes lead to incorrect
interpretations~\cite{Arina:2016cqj}. With the limits on the masses of
any hypothetical particle being pushed to higher and higher scales, the
theoretical uncertainties related with the new physics signals can moreover
sometimes be
quite severe, in particular if the associated scale and Bjorken-$x$ value lead
to probing the parton densities in a regime in which they are poorly
constrained~\cite{Frixione:2019fxg}.

On the other hand, it would be valuable to get estimates of the capabilities
of the future runs of the LHC with respect to a given signal, possibly on the
basis of the interpretation of the results of existing analyses of current data.
Predictions in which the signal and the background are naively scaled up could
hence be useful to obtain an initial guidance on the reach of future collider
setups within new physics parameter spaces.

In this paper, we address the above mentioned issues by 
presenting an extension of the recasting capabilities of the
\madanalysis\ platform~\cite{Dumont:2014tja,Conte:2018vmg} so that signal
theoretical and systematics uncertainties could be included in the recasting
procedure. Moreover, we show how the reinterpretation results, with
uncertainties included, could be correctly extrapolated to different
luminosities to get insight on the sensitivity of the future LHC data on given
signals.

As an illustration of
these new features within  concrete cases,
we consider  several classes of widely used simplified models. We
first extract bounds on various model parameters from recent LHC results. Next,
we study how those constraints are expected to evolve with the upcoming
high-luminosity run of the LHC through a naive rescaling of the signal and
background predictions. In practice, we make use of the recasting capabilities
of \madanalysis\ and pay a special attention to the theoretical uncertainties.

We begin with a simplified model inspired by the Minimal Supersymmetric
Standard Model (MSSM) in which the SM is complemented by a gluino and a
neutralino, all other superpartners being assumed heavy and
decoupled~\cite{Alwall:2008ag,Alves:2011wf}.
Such a particle spectrum leads to a signature comprised of jets and
missing transverse energy originating from the gluino decays into an invisible
neutralino and quarks. We reinterpret the results of corresponding
ATLAS searches for the signal  in 36~fb$^{-1}$~\cite{Aaboud:2017vwy}
and 139~fb$^{-1}$~\cite{ATLAS:2019vcq} of LHC data. We
investigate the impact of the theory errors on the derived bounds at the nominal
luminosity of the search, and extrapolate the findings to estimate the outcome
of  similar searches analysing 300 and
3000~fb$^{-1}$ of LHC data. Secondly, we make use of these recent
LHC searches to perform an equivalent exercise in the context of a simplified
model in which the SM is extended by a single species of first generation
squarks and a neutralino~\cite{Alwall:2008ag,Alves:2011wf}. Such a spectrum also
leads to a new physics signature made of jets and missing transverse energy,
although the squark colour triplet nature yields a signal featuring a
smaller jet multiplicity. As the considered ATLAS study includes a large set of
signal regions each dedicated to a different jet multiplicity, it is sensitive
to this simplified model that has moreover not been covered the result
interpretations performed in the experimental publication.

As a last example, we study the phenomenology
of a simplified dark matter model in which a Dirac fermion dark matter candidate
couples to the SM via interactions with an $s$-channel spin-1
mediator~\cite{Backovic:2015soa,Abercrombie:2015wmb}. This model is known to be
reachable via standard LHC monojet and multijet plus missing transverse energy
searches for dark matter. We extract up-to-date bounds on the model
by reinterpreting the results of the ATLAS search of ref.~\cite{ATLAS:2019vcq}
that analyses the full Run~2 ATLAS dataset. This search includes signal regions
dedicated to both the monojet and the multijet plus missing energy signatures,
so that it consists in an excellent probe for dark matter models. We
focus on two specific configurations of our generic simplified models in which
the mediator couples with the same strength to the dark and SM sectors. In the
first case, we consider mediator couplings of a vector nature, whilst in the
second case, we focus on axial-vector mediator couplings. We investigate how the
bounds evolve with the luminosity for various dark matter and mediator masses
and the nature of the new physics couplings.

The rest of this paper is organised as follows. We discuss the details of the
recasting capabilities of \madanalysis\ in section~\ref{sec:recasting}, focusing
not only on the new features that have been implemented in the context of this
work, but also on how the code should be used for LHC recasting. We then apply it
to extracting gluino and neutralino mass limits in section~\ref{sec:GluProd} for
various luminosities of LHC data.  We analyse the squark/neutralino
simplified model in section~\ref{sec:squarks} and perform our dark matter
analysis in section~\ref{sec:dm}. We summarise our work and conclude in
section~\ref{sec:conclusion}.

\section{LHC recasting with \madanalysis}
\label{sec:recasting}

The \madanalysis\ package~\cite{Conte:2012fm,Conte:2014zja} is a framework
dedicated to new physics phenomenology. Whilst the first aim of the
pro\-gram\-me was to facilitate the design and the implementation of analyses
targeting a given collider signal of physics beyond the Standard Model, and
how to unravel it from the background, more recently it has been extended by LHC
reinterpretation capabilities~\cite{Dumont:2014tja,Conte:2018vmg}. This feature
allows the user to derive the sensitivity of the LHC to any collider signal
obtained by matching hard-scattering matrix elements with parton showers, based
on the ensemble of analyses that have implemented in the
\href{https://madanalysis.irmp.ucl.ac.be/wiki/PublicAnalysisDatabase}
{\madanalysis\ public analysis database} (PAD)~\cite{Dumont:2014tja}%
\footnote{See the webpage
\url{https://madanalysis.irmp.ucl.ac.be/wiki/PublicAnalysisDatabase}.}. For each
of these analyses, the code simulates the experimental strategies (which
includes both the simulation of the detector response and the selection) to
predict the number of signal events that should populate the analysis
signal regions. It then compares the results with both data and the SM
expectation, so that conclusive statements could be drawn. As in all recasting
codes relying on the same method~\cite{Drees:2013wra,Buckley:2010ar,Balazs:%
2017moi}, the uncertainty on the signal is ignored although it could be
relevant~\cite{Frixione:2019fxg}.

With the release of \madanalysis\ version v1.8, the user has now the possibility
to deal with various classes of signal uncertainties and to extrapolate any
reinterpretation result to higher luminosities. This section documents all these
new functionalities.
Section~\ref{sec:install} briefly summarises how to install \madanalysis, get
the code running and download a local copy of its public analysis database.
Section~\ref{sec:runma5} details how the code can be used to reinterpret
the results of a specific LHC analysis. A more extensive and longer version of
this information on \madanalysis\ installation and running procedures
can be found in ref.~\cite{Conte:2018vmg}. Section~\ref{sec:newma5}
is dedicated to the new methods that have been developed in the context of this
work, and which are available from \madanalysis\ version v1.8 onwards. We also
introduce in this section several new optional features that can be used
for the design of the analysis {\tt.info} files. One such file accompanies each
analysis of the database and contains information on the observation and the
SM expectation of the different analysis signal regions. In
section~\ref{sec:output}, we describe the corresponding modifications of the
output format relevant for a recasting run of \madanalysis.

\subsection{Prerequisites and installation}\label{sec:install}

\madanalysis\ is compatible with most recent {\sc Unix}-based operating systems,
and requires the GNU \gpp\ or {\sc CLang} compiler, a {\sc Python} 2.7
installation (or more recent, but not a {\sc Python}~3 one) and {\sc GMake}. In
order for the recasting functionalities to be enabled, the user must ensure that
the {\sc SciPy} library is present, as it allows for limit computations, and
that the \delphes\ package~\cite{deFavereau:2013fsa} is locally available within
the \madanalysis\ installation. The
latter, which requires the \rooot\ framework~\cite{Brun:1997pa} and the
\fastjet\ programme~\cite{Cacciari:2011ma}, is internally called by
\madanalysis\ to deal with the simulation of the response of the LHC detectors
and to reconstruct the events. Moreover, reading compressed event files can only
be performed if the {\sc Zlib} library is available.

The latest version of \madanalysis\ can be downloaded from
\href{https://launchpad.net/madanalysis5}{\sc LaunchPad}%
\footnote{See the webpage
\url{https://launchpad.net/madanalysis5}.},
where it is provided as a tarball named
{\tt ma5\_v<xxx>.tgz},  that contains all \madanalysis\ source files ({\tt <xxx>}
standing for the version number). After unpacking the tarball, the code can be
started by issuing in a shell
\begin{verbatim}
  ./bin/ma5 -R
\end{verbatim}
where the {\tt -R} options enforces the {\tt reco} mode of \madanalysis, that is
relevant for LHC recasting. The programme begins with checking the presence of
all mandatory packages and determining which of the optional packages are
available. The \madanalysis\ command-line interface is then initialised and the
user is prompted to type in commands.

In the case where any of the {\sc Zlib} or \delphes\ package would not be found
by \madanalysis, they can be installed locally by typing, directly in the
\madanalysis\ interpreter,
\begin{verbatim}
  install zlib
  install delphes
\end{verbatim}
Whilst \rooot\ can in principle be installed similarly, we recommend the user to
handle this manually, following the instructions available on the
\href{https://root.cern.ch/}{\rooot\ website}\footnote{See the
webpage \url{https://root.cern.ch}.}
Furthermore, all existing and validated recast LHC analyses in the \madanalysis\
framework can be locally downloaded by typing in,
\begin{verbatim}
  install PAD
  install PADForMA5tune
\end{verbatim}
The second command triggers the installation of older implemented analyses, that
requires a (now disfavoured) {\tt MA5tune} version of \delphes. The latter can
be installed by typing, in the \madanalysis\ shell,
\begin{verbatim}
  install delphesForMA5tune
\end{verbatim}

\subsection{Recasting LHC analyses with \madanalysis}\label{sec:runma5}

In this section, we rely on a generic example in which a user aims to estimate
the sensitivity of a specific LHC analysis to a given signal with \madanalysis.
The analysis consists of one of the analyses available from the PAD and the
signal is described by simulated events collected into a file that we call
\verb+events.hepmc.gz+. Such an event file includes the simulation of the
considered hard-scattering process matched with parton showers, as well as the
hadronisation of the final-state partons present in each of the showered events.

As mentioned above, \madanalysis\ has to be star\-ted in the {\tt reco} mode,
\begin{verbatim}
  ./bin/ma5 -R
\end{verbatim}
In a first step, the recasting mode of the programme has to be enabled and the
event file, physically located at \verb+<path-to-events.hepmc.gz>+ on the user
system, has to be im\-por\-ted. This is achieved by issuing the commands
\begin{verbatim}
  set main.recast = on
  import <path-to-events.hepmc.gz> as <label>
\end{verbatim}
The second command defines a dataset identified by the label
\verb+<label>+ that here solely includes the imported sample. Several event
files can be imported and collected either under a unique dataset (by using the
same \verb+<label>+ for each call to the \verb+import+ command) or split into
different datasets (by employing different labels). When studying the signal
under consideration, \madanalysis\ will run over all defined datasets and
imported event files.

In addition, the user can activate the storage of the \rooot\ file(s) generated
by \delphes\ by issuing the command,
\begin{verbatim}
  set main.recast.store_root = <status>
\end{verbatim}
where {\tt <status>} can take the \verb+True+ or \verb+False+ value, and
directly provide a predefined recasting card (available on the system at {\tt
<path-to-a-card>}), through
\begin{verbatim}
  set main.recast.card_path = <path-to-a-card>
\end{verbatim}
In the case where no card is provided, \madanalysis\ creates a consistent
new card with one entry for each of the available analyses. Such an entry is of
the form
\begin{verbatim}
  <tag> <type> <switch> <detector> # <comment>
\end{verbatim}
The \verb+<tag>+ label corresponds to the filename of the C++ code associated
with the considered analysis (located in the
\verb+Build/SampleAnalyzer/User/Analyzer+ subdirectory of the PAD installation
in \verb+tools/PAD+), the \verb+<type>+ label indicates whether the
PADForMA5tune (\verb+v1.1+) or PAD (\verb+v1.2+) recasting infrastructure should
be used and the \verb+<switch>+ tag (to be set to \verb+on+ or \verb+off+)
drives whether the analysis has to be recast. The name of the \delphes\
card to use (see the \verb+Input/Cards+ subdirectory of the PAD installation)
is passed as \verb+<detector>+, and \verb+<comment>+ consists of an optional
comment (usually briefly describing the analysis).

The run is finally started by typing in the interpreter,
\begin{verbatim}
  submit
\end{verbatim}
Firstly, \madanalysis\ simulates the detector impact on the input events,
for each of the necessary \delphes\ cards according to the analyses that have
been switched on in the recasting card. Next, the code derives how the different
signal regions are populated by the signal events and finally computes, by means
of the CL$_s$ prescription\cite{Read:2002hq}, the corresponding exclusion
limits, signal region by signal region. This is achieved by a comparison of the
results with the information on the SM background and data available from
the different \verb+info+ files shipped with the PAD.

The output information is collected into a folder named \verb+ANALYSIS_X+, where
\verb+X+ stands for the next available positive integer (in terms of
non-existing directories). On top of basic details about the run itself, this
folder contains the recasting results that are located in the
\verb+ANALYSIS_X/Output+
folder. The latter includes the \verb+CLs_output_summary.dat+ file that
concisely summarises all the results of the run. A more extensive version of
these results can be found in the set of subfolders named after the labels of
the imported datasets. The \verb+CLs_output_summary.dat+ file contains one line
for each signal region of each reinterpreted analysis, and this for each of the
datasets under consideration. Each of these lines follows the format
\begin{verbatim}
  <set> <tag> <SR> <exp> <obs> || <eff> <stat>
\end{verbatim}
where the \verb+<set>+ and \verb+<tag>+ elements respectively consist in the
names of the dataset and analysis relevant for the considered line of the output
file. The \verb+<SR>+ entry relates to one of the analysis signal regions, the
exact name being the one defined in the analysis C++ source code. The
\verb+<exp>+ and \verb+<obs>+ quantities are the expected and observed
cross-section values for which the signal modelled by the events stored within
the dataset \verb+<set>+ is excluded by the signal region \verb+<SR>+ of the
analysis \verb+<tag>+ at the 95\% confidence level. In the former case, the code
makes use of the SM expectation to predict the number of events populating the
signal region \verb+<SR>+, whilst in the latter case, data is used. Finally, the
\verb+<eff>+ and \verb+<stat>+ entries respectively refer to the corresponding
selection efficiency and the associated statistical error.

The user has the option to specify the cross section corresponding to the
investigated signal by issuing, in the \madanalysis\ interpreter,
\begin{verbatim}
  set <label>.xsection = <value>
\end{verbatim}
prior to the call to the \verb+submit+ command.
Following this syntax, \verb+<label>+ stands for one of the labels of the
considered datasets and \verb+<value>+ for the associated cross-section value,
in pb. In this case, the confidence level at which the analysed signal is
excluded is included in the output summary file (before the double vertical
line).

The \verb+Output+ folder additionally contains a specific subfolder for each of
the defined datasets. Such a directory contains a file named
\verb+CLs_output.dat+ that includes the same information as in the
\verb+CLs_output_summary.dat+ file, following the same syntax, but restricted to
a specific dataset. A second file encoded into the SAF
format~\cite{Conte:2012fm} and named \verb+<label>.saf+ (\verb+<label>+
being the dataset name) contains general information on the dataset organised
according to an XML-like structure. The latter relies on three classes of
elements, namely
\verb+<SampleGlobalInfo>+, \verb+<FileInfo>+ and \verb+<SampleDetailedInfo>+.
The first of the\-se contains global information on the dataset, such as its
cross section (\verb+xsec+), the associated error (\verb+xsec_err+), the number
of events (\verb+nev+) or the sum of the positive and negative event weights
(\verb+sum_w++ and \verb+sum_w-+). The corresponding entry in the output file
would read
\begin{verbatim}
 <SampleGlobalInfo>
    # xsec  xsec_error  nev    sum_w+  sum_w-
      ...      ...      ...    ...     ...
 </SampleGlobalInfo>
\end{verbatim}
where the numerical values have been omitted for clarity. The \verb+<FileInfo>+
element sequentially provides the paths to the different event files included in
the dataset, while detailed information on each file is provided within the
\verb+<SampleDetailedInfo>+ XML root element, in a similar manner as for the
sample global information (with one line for each file).

Furthermore, the dataset output directory includes a \verb+RecoEvents+
folder dedicated to the storage of \delphes\ output files (one file for each
considered detector parameterisation), provided that the corresponding option
has been turned on (see above), as well as one folder for each of the recast
analyses. Each of these folders contains one SAF file listing all signal regions
implemented in the associated analysis, as well as two subfolders
\verb+Cutflows+ and \verb+Histograms+. The former includes one SAF file for each
signal region, and the latter a single file named
\verb+histos.saf+.

A cutflow is organised through XML-like elements, \verb+<InitialCounter>+ and
\verb+<Counter>+ being used for the initial number of events and the results of
each selection cut respectively. As depicted by the example below, in which all
numbers have been omitted for clarity,
\begin{verbatim}
  <Counter>
  "my_cut_name"   # 1st cut
  ....    ....    # nentries
  ....    ....    # sum of weights
  ....    ....    # sum of weights^2
  </Counter>
\end{verbatim}
any of such elements includes a cut name as defined in the analysis C++ file
(first line), the number of events passing the cut (second line), the weighted
number of events passing the cut (third line) and the sum of the squared weights
of all events passing the cut (last line). Moreover, the first (second) column
refers to the positively-weighted (negatively-weighted) events only.

Histograms are all collected into the file \verb+histos.saf+, that is also
organised according to an XML-like structure relying on several \verb+<Histo>+
elements. Each of these corresponds to one of the histograms implemented in the
analysis. A \verb+<Histo>+ element includes the definition of the histogram
(provided within the \verb+<Description>+ element), general statistics (as part
of the \verb+<Statistics>+ element) and the histogram data itself (within the
\verb+<Data>+ element). The description of a histogram schematically reads
\begin{verbatim}
  <Description>
    "name"
    # nbins   xmin           xmax
      ..      ...            ...
    # Defined regions
      ...   # Region nr. 1
      ...   # Region nr. 2
  </Description>
\end{verbatim}
and is self-explanatory, all numbers having been replaced by dots. This moreover
shows that a given histogram can be associated with several signal
regions, provided they are indistinguishable at the moment the histogram is
filled. Statistics are typically given as
\begin{verbatim}
 <Statistics>
  ...  ...  # nevents
  ...  ...  # sum of event-weights over events
  ...  ...  # nentries
  ...  ...  # sum of event-weights over entries
  ...  ...  # sum weights^2
  ...  ...  # sum value*weight
  ...  ...  # sum value^2*weight
 </Statistics>
\end{verbatim}
which include information about the number of entries, the weighted number of
entries, the variance, \etc\ Moreover, the contributions of the
positively-weighted and ne\-ga\-ti\-ve\-ly-weighted events are again split and
provided with\-in the first and second column respectively. The values of each
bin are finally available from the \verb+<Data>+ element,
\begin{verbatim}
  <Data>
    ...  ...    # underflow
    ...  ...    # bin 1 / 15
       .
       .
       .
    ...  ...    # bin 15 / 15
    ...  ...    # overflow
  </Data>
\end{verbatim}
where all bin values are omitted and the two columns respectively refer to
events with positive (first column) and negative (second column) weights. The
underflow and overflow bins are also included.

To close this section, we detail below how limits on a given signal are derived
by \madanalysis, using the CL$_s$ prescription. The output file generated by the
code contains three numbers associated with those limits, the expected and
observed cross sections excluded at the 95\% confidence level,
$\sigma_{95}^{\rm exp}$ and $\sigma_{95}^{\rm obs}$, as well as the confidence
level at which the input signal is excluded. Those numbers are extracted on the
basis of the information available from the \verb+.info+ file, shipped with each
recast analysis and that contains, for each signal region, the number of
expected SM events $n_b$, the associated error $\Delta n_b$ and the observed
number of events populating the signal region $n_{\rm obs}$. As said above,
starting from the input event file, \madanalysis\ simulates the response of the
LHC detector, applies the analysis selection, and estimates how the different
signal regions are populated. In this way, for each signal region, the number of
signal events $n_s$ is known.

This enables the computation of the background-only and signal-plus-background
probabilities $p_{\rm b}$ and $p_{{\rm b} + {\rm s}}$ and to further derive the
related CL$_s$ exclusion. In practice, the code considers a number of toy
experiments (the default being 100000 that can be changed by issuing, in the
\madanalysis\ interpreter and before the call to the \verb+submit+ method,
\begin{verbatim}
  set main.recast.CLs_numofexps = <value>
\end{verbatim}
where \verb+<value>+ stands for the desired number of toy experiments. For each
toy experiment, the expected number of background events $N_{\rm b}$ is randomly
chosen assuming that its distribution is Gaussian, with a mean $n_{\rm b}$ and a
width $\Delta n_{\rm b}$. The corresponding probability density thus reads
\be
  f(N_b|n_b,\Delta n_b) =
    \frac{\exp\left\{-\frac{(N_b - n_b)^2}{2\Delta n_b^2}\right\}}
    {\Delta n_b\sqrt{2\pi}} \ .
\ee
Imposing $N_{\rm b} \geq 0$, the actual number of background events
$\hat N_{\rm b}$ is randomly generated from the Poisson distribution
\be
  f(\hat{N}_b|N_b) = \frac{N^{\hat{N}_b}_be^{-N_b}}{\hat{N}_b!}\ .
\ee
Accounting for the observation of $n_{\rm obs}$ events, $p_{\rm b}$ is defined
as the percentile of score associated with $ \hat N_{\rm b} \leq n_{\rm obs}$,
which consists in the probability for the background to fluctuate as low as
$n_{\rm obs}$.

The signal-plus-background probability $p_{{\rm b} + {\rm s}}$ is computed
similarly, assuming that the actual number of signal-plus-background events
$\hat N_{\rm b}+\hat N_{\rm s}$ follows a Poisson distribution of parameter
$n_{\rm s}+N_{\rm b}$ (after imposing this time that $N_{\rm b} + n_{\rm s}>0$).
The resulting CL$_s$ exclusion is then derived as
\be
  {\rm CL}_s = {\rm max}\Big(0, 1-\frac{p_{{\rm b}+{\rm s}}}{p_{\rm b}}\Big) \ .
\ee
and $\sigma_{95}^{\rm obs}$ is calculated as above in a case where the number of
signal events $n_s$ is kept free. From the (derived) knowledge of the analysis
selection efficiencies, \madanalysis\ can extract the upper allowed cross
section value for which the signal is not excluded, \ie~$\sigma_{95}^{\rm obs}$.
The expected cross section excluded at the 95\% confidence level,
$\sigma_{95}^{\rm exp}$, is obtained by replacing $n_{\rm obs}$ by $n_{\rm b}$
in the above calculations.

\subsection{Including signal uncertainties and extrapolation to higher
  luminosities}\label{sec:newma5}
In the procedure described in the previous section, any error on the signal is
ignored, both concerning the usual theory uncertainties (scale variations,
parton densities) and the systematics, mostly stemming from more experimental
aspects. In particular, with the constantly growing mass bounds on hypothetical
new particles, the scale entering the relevant hard-scattering processes is
larger and larger, so that theoretical errors could start to impact the derived
limits in an important and non-negligible manner.

Starting from version v1.8 onwards, \madanalysis\ offers the user a way to
account for both the theoretical and systematical errors on the signal when a
limit calculation is performed. The scale and parton density (PDF) uncertainties
can be entered, within the \madanalysis\ interpreter, similarly to the cross
section associated with a given dataset (see section~\ref{sec:runma5}),
\begin{verbatim}
  set <label>.xsection        = <xsec_val>
  set <label>.scale_variation = <scale>
  set <label>.pdf_variation   = <pdf>
\end{verbatim}
where \verb+<label>+ stands for the label defining the signal dataset. In this
case, the signal cross section $\sigma_s$ is provided through the
\verb+xsection+ attribute of the dataset, as described in the previous section,
while the scale and parton density uncertainties $\Delta\sigma_{\rm scales}$ and
$\Delta\sigma_{\rm PDF}$ are given through the \verb+scale_variation+ and
\verb+pdf_variation+ attributes. The errors are symmetric with respect to the
central value $\sigma_s$, and their value (given by \verb+<scale>+ and
\verb+<pdf>+ in the above example) must be inputted as the absolute values of
the relative errors on the cross section (\ie~as positive floating-point
numbers). Asymmetric errors can also be provided, the upper and lower
uncertainties being independently fixed by issuing, in the \madanalysis\
interpreter,
\begin{verbatim}
  set <label>.scale_up_variation   = <scale_up>
  set <label>.scale_down_variation = <scale_dn>
  set <label>.pdf_up_variation     = <pdf_up>
  set <label>.pdf_down_variation   = <pdf_dn>
\end{verbatim}
Each error is again provided as a positive floating-point number and refers to
the relative error on the cross section, in absolute value. On top of the
computation of the confidence level at which the signal is excluded,
\madanalysis\ additionally calculates the CL$_s$ variation band associated with
the scale uncertainties, as well as with the total theory uncertainties where
both the scale and PDF contributions to the total error are added linearly.
Such a behaviour can however be modified by issuing, in the interpreter
\begin{verbatim}
  set main.recast.THerror_combination = <value>
\end{verbatim}
where \verb+<value>+ can be set either to \verb+quadratic+ (the theory errors
are added quadratically) or \verb+linear+ (default, the theory errors are added
linearly). The CL$_s$ band is then derived by allowing the signal cross section
to vary within its error band, deriving the associated spread on
$p_{\rm{b}+\rm{s}}$.

The user can also specify one or more values for the level of systematics on the
signal. This is achieved by issuing, in the command line interface,
\begin{verbatim}
  set main.recast.add.systematics = <syst>
\end{verbatim}
This command can be reissued as many times as needed, \madanalysis\ taking care
of the limit calculation for each entered value independently. The level of
systematics (\verb+<syst>+) has to be given either as a floating-point number
lying in the $[0,1]$ range, or as a pair of floating-point numbers lying in the
same interval. In the former case, the error is symmetric with respect to the
central value $\sigma_s$, whilst in the latter case, it is asymmetric with the
first value being associated with the upper error and the second one with the
lower error.

In addition, we have also extended the code so that naive
extrapolations for a different luminosity ${\cal L}_{\rm new}$ could be
performed. This is achieved by typing, in the interpreter,
\begin{verbatim}
  set main.recast.add.extrapolated_luminosity \
    = <lumi>
\end{verbatim}
Once again, the user has the possibility to reissue the command several times,
so that the extrapolation will be performed for each luminosity \verb+<lumi>+
independently (where the value has to be provided in fb$^{-1}$). Those
extrapolations assume that the signal and background selection efficiencies of a
given region in a specific analysis are identical to those corresponding to the
reference luminosity ${\cal L}_0$ initially considered. In this framework, the
extrapolated number of background events $n_b^{\rm new}$ is related to $n_b$ (the
number of background events expected for the reference luminosity ${\cal L}_0$) as
\be
  n_b^{\rm new} =  n_b\ \frac{\mathcal{L}_{\rm new}}{\mathcal{L}_0}  \ ,
\ee
 that we assume equal to the extrapolated number of observed events,
\be
  n_{\rm obs}^{\rm new} = n_b^{\rm new} \ .
\ee
On the other hand, the associated uncertainties, $\Delta n_b^{\rm new}$, are
derived from the relation
\be
 \Delta n_b^{\rm new} =
    \Delta_{b, {\rm syst}} \frac{{\cal L}_{\rm new}}{{\cal L}_0} \oplus
    \Delta_{b, {\rm stat}} \sqrt{\frac{{\cal L}_{\rm new}}{{\cal L}_0}} \ ,
\label{eq:bgd_xtp}\ee
where the statistics and systematics components are added in quadrature. The
systematics are extrapolated linearly, whilst the statistical uncertainties
assume that the event counts follow a Poisson distribution. Such an
extrapolation of the background error requires an access to the details of the
background uncertainties. This is however not achievable within the XML
\verb+info+ file format dedicated to the transfer of the background and data
information to \madanalysis~\cite{Dumont:2014tja}. We therefore introduce two
new XML elements to this format, namely \verb+deltanb_stat+ and
\verb+deltanb_syst+. These offer the user the option to implement his/her
\verb+info+ file by either providing a unique combined value for the
uncertainties (via the standard \verb+deltanb+ XML element) or by splitting them
into their statistical and systematical components (via a joint use of the new
\verb+deltanb_stat+ and \verb+deltanb_syst+ XML elements). In this way, a region
element could be either implemented according to the old syntax, as in the
schematic example below (with all numbers omitted),
\begin{verbatim}
  <region type="signal" id="Region name">
    <nobs>    ... </nobs>
    <nb>      ... </nb>
    <deltanb> ... </deltanb>
  </region>
\end{verbatim}
or following the new syntax, which would then read
\begin{verbatim}
  <region type="signal" id="Region name">
    <nobs>         ... </nobs>
    <nb>           ... </nb>
    <deltanb_stat> ... </deltanb_stat>
    <deltanb_syst> ... </deltanb_syst>
  </region>
\end{verbatim}
Whilst the usage of the new syntax is encouraged, this  new
possibility for embedding the error information strongly
depends on how the background uncertainties are provided in the experimental
analysis notes. For this reason, as well as for backward-compatibility,
\madanalysis\ supports both choices. If only a global error is provided, the
user can freely choose how to scale the error (linearly or in a Poisson way),
by typing in the interpreter,
\begin{verbatim}
 set main.recast.error_extrapolation = <value>
\end{verbatim}
where \verb+<value>+ has to be set either to \verb+linear+ or to \verb+sqrt+.
The user has also the choice to use a single floating-point number
for the \verb+<value>+ parameter. In this case, the relative
error on the number of background events at the new luminosity,
$\Delta n_b^{\rm new}/n_b^{\rm new}$, is taken equal to this number. Finally, the
user can provide a comma-separated pair of floating-point numbers $\kappa_1$ and
$\kappa_2$, as in 
\begin{verbatim}
 set main.recast.error_extrapolation = <k1>,<k2>
\end{verbatim}
The background error is here defined by
\be
  \bigg[\frac{\Delta n_b^{\rm new}}{n_b^{\rm new}}\bigg]^2 =
     \kappa_1^2 + \frac{\kappa_2^2}{n_b} \ ,
\ee
where the two values provided by the user respectively control the systematical
component of the uncertainties (\verb+<k1>+, $\kappa_1$) and the statistical one
(\verb+<k2>+, $\kappa_2$).
Finally, all extrapolations are based on expectations and not on observations,
so that $n_{\rm obs}$ will be effectively replaced by the corresponding SM
expectation $n_{\rm b}$.

\subsection{Output format}\label{sec:output}
\madanalysis\ propagates the information on the impact of the uncertainties all
through the output file, which is then written in a format slightly extending
the one presented in section~\ref{sec:runma5}. Starting with the summary file
\verb+CLs_output_summary.dat+, each line (corresponding to a given signal region
of a given analysis) is now followed by information schematically written as
\begin{verbatim}
  Scale var. band                [..., ...]
  TH   error band                [..., ...]
  +<lvl_up>%, -<lvl_dn>% syst    [..., ...]
\end{verbatim}
The uncertainties on the
exclusion stemming from scale variations are given in the first line, which is
trivially omitted if the corresponding information on the signal cross section
is not provided by the user. In the second line, \madanalysis\ adds either
quadratically or linearly (according to the choice of the user) all theory
errors, such a line being written only if at least one source of theory
uncertainties is provided by the user. Finally, if the user inputted one or more
options for the level of systematics, \madanalysis\ computes the band resulting
from the combination of all errors and writes it into the output file (one line
for each choice of level of systematics). In the above snippet, the user fixed
an asymmetric level of systematics (for the sake of the example) indicated by
the \verb+<lvl_up>+ and \verb+<lvl_dn>+ tags.

In cases where the band would have a vanishing size, the uncertainty
information is not written to the output file. This could be due either to
negligibly small uncertainties, to the fact that for the considered
region, the signal is excluded regardless the level of systematics (at the 100\%
confidence level), or to the region not targeting the signal at all (the
corresponding selection efficiency being close to zero).

The \verb+CLs_output.dat+ dataset-specific files present in the output
subdirectory associated with each imported dataset all contain similar
modifications. In case of extrapolations to different luminosities, copies of
this file named \verb+CLs_output_lumi_<lumi>.dat+ are provided for each desired
luminosity \verb+<lumi>+.

\section{Gluino and neutralino mass limits}\label{sec:GluProd}
\begin{figure}
  \centering
  \includegraphics[width=0.70\columnwidth]{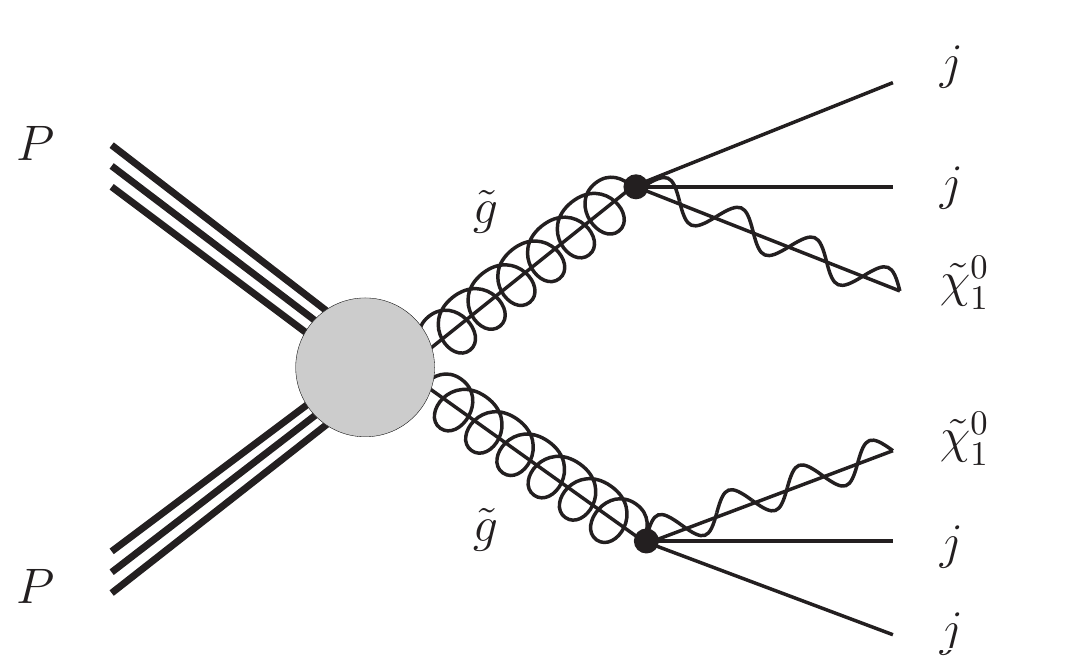}
  \caption{\it Generic Feynman diagram associated with the production and decay
    of a pair of gluinos in the considered MSSM-inspired gluino
    simplified model. The figure has been produced with the help of the
    {\sc JaxoDraw} package~\cite{Binosi:2008ig}.
  \label{fig:diag}}
\end{figure}

To illustrate the usage of the new functionalities of \madanalysis\ introduced
in the previous section, we perform several calculations in the context of a
simplified model inspired by the MSSM.
In this framework, all superpartners are heavy and decoupled, with the exception
of the gluino $\tilde g$ and the lightest neutralino $\tilde\chi_1^0$, taken to
be bino-like. Any given benchmark is thus defined by two parameters, namely
the gluino and the neutralino masses $m_{\tilde g}$ and $m_{\tilde\chi_1^0}$.
Such a new physics setup can typically manifest itself at the LHC through a
signature made of a large hadronic activity and missing transverse energy. As
shown by the schematic Feynman diagram of figure~\ref{fig:diag}, such a
signature originates from the production of a pair of gluinos, each of them
promptly decaying into two jets and a neutralino (via virtual squark
contributions).

We study the sensitivity of the LHC and its higher-luminosity upgrades to this
signal by analysing state-of-the-art Monte Carlo simulations
a\-chie\-ved by means of the \madgraph\ framework (version 2.6.6)~\cite{%
Alwall:2014hca}, using the MSSM-NLO model implementation developed in
ref.~\cite{Frixione:2019fxg}. Hard-scattering  matrix elements are generated at
the next-to-leading-order (NLO) accuracy in QCD and convoluted with the NLO set
of NNPDF~3.0 parton densities~\cite{Ball:2014uwa}, as provided by the \lhapdf\
interface~\cite{Buckley:2014ana}. The gluino leading-order (LO) decays are
handled with the
\madspin~\cite{Artoisenet:2012st} and \madwidth~\cite{Alwall:2014bza}
packages. The resulting NLO matrix elements are then matched with \pythia\
parton showers and hadronisation (version 8.240)~\cite{Sjostrand:2014zea},
following the MC@NLO method~\cite{Frixione:2002ik}. Our predictions include
theoretical uncertainties stemming from the independent variations of the
renormalisation and factorisation scales by a factor of two up and down
relatively to the central scale, taken as half the sum of the transverse masses
of the final-state particles, as well as from the parton densities extracted
following the recommendations of ref.~\cite{Demartin:2010er}.

\begin{figure}
  \centering
  \includegraphics[width=0.97\columnwidth]{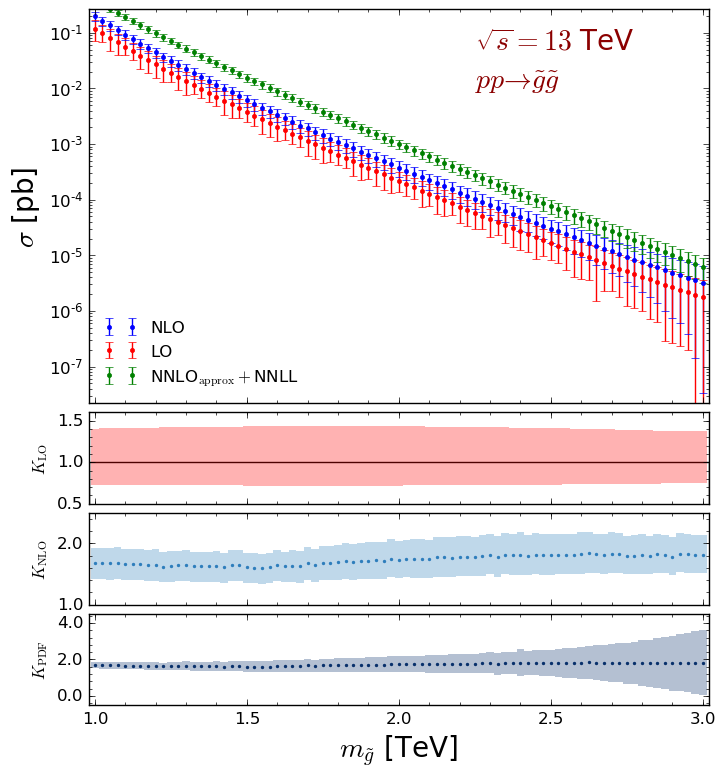}
  \caption{\it Total LO (red), NLO (blue) and
    NNLO$_{\rm approx}$+NNLL (green) cross sections (upper panel) and
    $K$-factors (three lower panels, where the results are normalised to the
    LO central value) for gluino pair-production, at a centre-of-mass energy of
    $\sqrt{s}=13$~TeV. In the upper panel, the error bands correspond to the
    quadratic sum of the scale and PDF uncertainties, whilst in the second and
    third panels, respectively, they  refer to the scale
    uncertainties on the LO and NLO predictions. The last panel focuses on the
    PDF errors.
  \label{fig:gluprod}}
\end{figure}

In the upper panel of figure~\ref{fig:gluprod}, we present the total LO
 (red) and NLO  (blue)
gluino pair-production cross section for gluino masses ranging from 1 to 3~TeV,
the error bars being associated with the quadratic sum of the scale and PDF
uncertainties. The cross section central value is found to vary within the
$100-0.001$~fb range when the gluino mass varies from 1 to 3~TeV, so that at
least tens of gluino events could be expected even for a very heavy gluino
benchmark at a high-luminosity upgrade of the LHC.  We compare our
predictions with the total rates traditionally employed by the ATLAS and CMS
collaborations to extract gluino limits, as documented by the LHC Supersymmetry
Cross Section Working Group~\cite{Borschensky:2014cia}.  Hence we include, in the
first panel of figure~\ref{fig:gluprod}, total gluino-pair production cross
sections matching approximate fixed-order results at next-to-next-to-leading
order and threshold-re\-sum\-med predictions at the next-to-next-to-leading
logarithmic accuracy (NNLO$_{\rm approx}$ + NNLL, in green). Following the
PDF4LHC recommendations, those more accurate NNLO$_{\rm approx}$+NNLL
predictions are obtai\-ned by convoluting the partonic cross section with a
combination of NLO CTEQ6.6M~\cite{Nadolsky:2008zw} and
MSTW2008~\cite{Martin:2009iq} densities. This choice, together with the impact
of the higher-order corrections, leads to NNLO$_{\rm approx}$+NNLL results
greater than our NLO predictions by a factor of about 2. While in the following
we use NLO-accurate total rates (as the latter exist for any new physics model
through a joint use of {\sc FeynRules}~\cite{Alloul:2013bka}, {\sc NLOCT}~\cite{
Degrande:2014vpa} and \madgraph), we evaluate the impact of higher-order
corrections whenever the relevant calculations exist, \ie\ in this section and
section~\ref{sec:squarks}.

With the second and third panels
of the figure, we emphasise the significant reduction of the scale uncertainties
at NLO by depicting the LO and NLO scale uncertainty bands respectively, the
$K_{\rm LO}$ and $K_{\rm NLO}$ quantities, presented in the two subfigures,
these being
the LO and NLO cross sections normalised to the LO central value. Such better
control in the theoretical predictions is one of the main motivations
for relying on NLO simulations instead of on LO ones. In the lower panel of
figure~\ref{fig:gluprod}, we focus on the PDF uncertainties associated with the
total rates and present the $K_{\rm PDF}$ quantity where the NLO result (with
its PDF error band) is again shown relatively to the LO central result. We omit
the corresponding LO curve, as it is similar to the NLO one, the same PDF set
being used both at LO and NLO in order to avoid having to deal with the
poor-quality LO NNPDF 3.0 fit~\cite{Ball:2014uwa}. Whilst the uncertainties are
under good control
over most of the probed mass range, the poor PDF constraints in the large
Bjorken-$x$ regime lead to predictions plagued by sizeable uncertainties
for gluino heavier than about $2.6-2.7$~TeV. Finally, our results show
that the NLO $K$-factor $K_{\rm NLO}$ is of about $1.6-1.7$, a typical value for
a strong supersymmetric production process, and features a significant gluino
mass dependence. The latter originates from the quark-antiquark contributions to
the cross section that become relatively larger with respect to the gluon fusion
ones with increasing Bjorken-$x$ values~\cite{Beenakker:1996ch}.

\begin{figure}
  \centering
  \includegraphics[width=0.97\columnwidth]{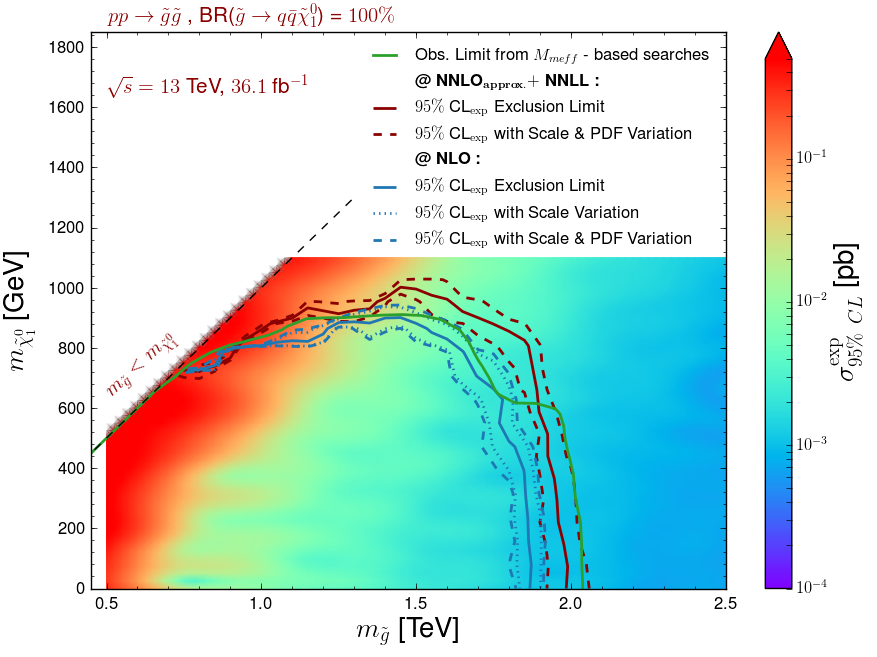}
  \caption{\it Constraints on the gluino-neutralino simplified model under
    consideration, represented as 95\% confidence level exclusion contours in
    the $(m_{\tilde g}, m_{\tilde\chi_1^0})$ plane. We compare the exclusion
    obtained with the ATLAS-SUSY-2016-07 reimplementation in the \madanalysis\
    framework~\cite{Chalons:AtlasSusy1607}  when normalising the
    signal to NLO (blue) and to NNLO$_{\rm approx}$+NNLL (red) with the
    official ATLAS results, extracted using the $M_{\rm eff}$ signal regions
    only~\cite{Aaboud:2017vwy} (solid green). Moreover, we include the
    uncertainty band on the \madanalysis\ results as originating from scale
    uncertainties (dotted) and from the
    quadratic combination of the scale and PDF uncertainties (dashed). The
    colour scheme represents the cross section value excluded
    at the 95\% confidence level for each mass configuration.
  \label{fig:exclusion_36}}
\end{figure}

We then predict, for several $(m_{\tilde g},m_{\tilde\chi_1^0})$ configurations,
how the signal events would populate the different signal regions of the
ATLAS-SUSY-2016-07 search for supersymmetry~\cite{Aaboud:2017vwy}. In practice,
we use the corresponding recast analysis as implemented in the \madanalysis\
public database~\cite{Chalons:AtlasSusy1607},
together with the appropriate \delphes\
configuration for the simulation of the detector response. In this analysis,
the ATLAS collaboration investigates the potential of a signature featuring
multiple jets and missing transverse energy through two approaches. The first
one relies on the so-called effective mass $M_{\rm eff}(N)$, a variable defined as
the scalar sum of the transverse momenta of the $N$ leading jets and the missing
transverse energy. The second one is based on the recursive jigsaw
reconstruction technique~\cite{Jackson:2017gcy}. Whilst all $M_{\rm eff}$-based
signal regions have been implemented in \madanalysis, the recursive jigsaw
reconstruction ones have been ignored due to the lack of information allowing
for their proper recasting. They are thus ignored in the following study as
well.

Our results are presented in figure~\ref{fig:exclusion_36} in the form of
exclusion contours in the $(m_{\tilde g},m_{\tilde\chi_1^0})$ mass plane, to
which we supplement the values of the signal cross section that are excluded at
the 95\% confidence level through a colour code. The exclusion
contours and excluded cross sections at the 95\% confidence level are extracted
by means of Gaussian process regression with a conservative amount of data as
implemented in the {\sc Excursion} package~\cite{lukas_heinrich_2018_1634428}.

We compare our predictions (the solid  blue line),
obtained with the setup described above, with the official ATLAS limits (the
green line) as originating from the
$M_{\rm eff}$-based signal region yielding the best expectation. ATLAS
simulations are based on calculations at the LO accuracy in which
samples of events describing final states featuring up to two extra jets are
merged~\cite{Lonnblad:2011xx}. Moreover, the ATLAS results are normalised to NLO
cross sections matched with threshold resummation at the next-to-leading
logarithmic accuracy (NLO+NLL)~\cite{Borschensky:2014cia}.
The ATLAS setup therefore differs from ours both at the level of the differential distributions, as we model
the properties of the second radiation jet solely at the level of the parton
showers, and at the level of the total rates that are evaluated at the NLO
matched with parton showers (NLO+PS) accuracy. This consequently results in
\madanalysis\ limits slightly weaker than the ATLAS ones by about 10\%,
especially in the light neutralino mass regime.
\begin{figure*}
  \centering
  \includegraphics[width=0.48\textwidth]{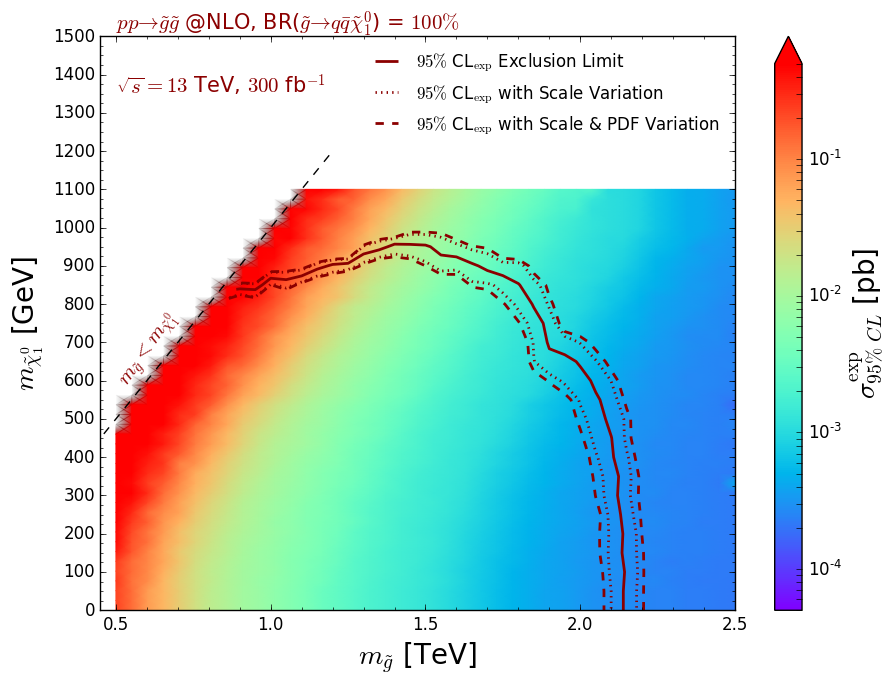}
  \hspace{5mm}
  \includegraphics[width=0.48\textwidth]{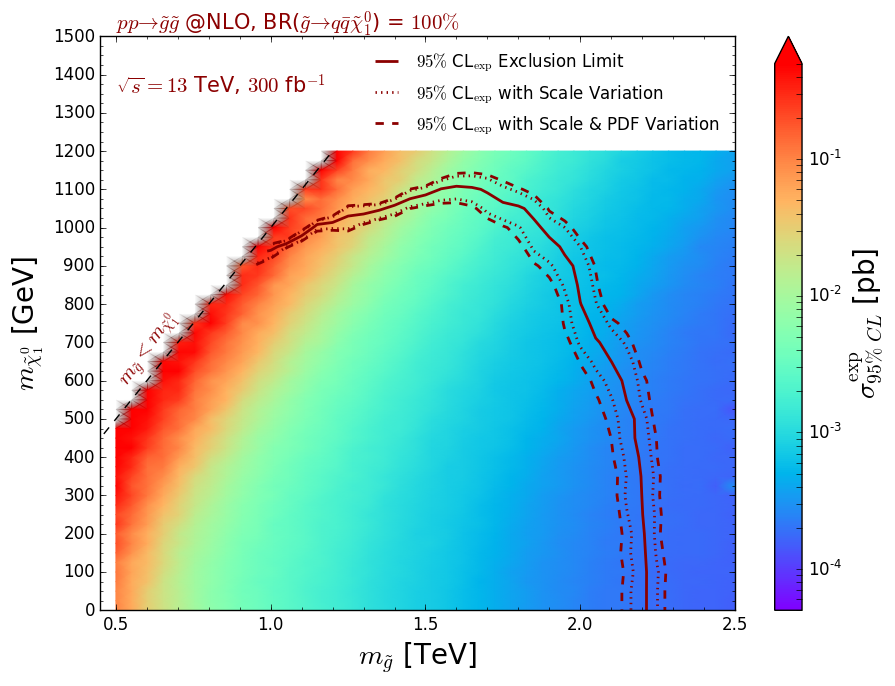}\\
  \includegraphics[width=0.48\textwidth]{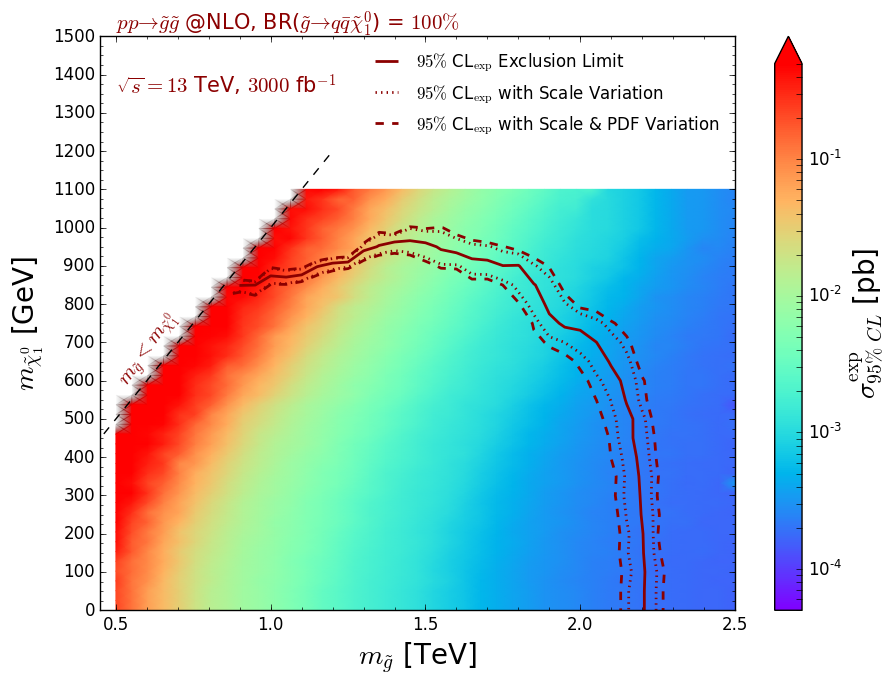}
  \hspace{5mm}
  \includegraphics[width=0.48\textwidth]{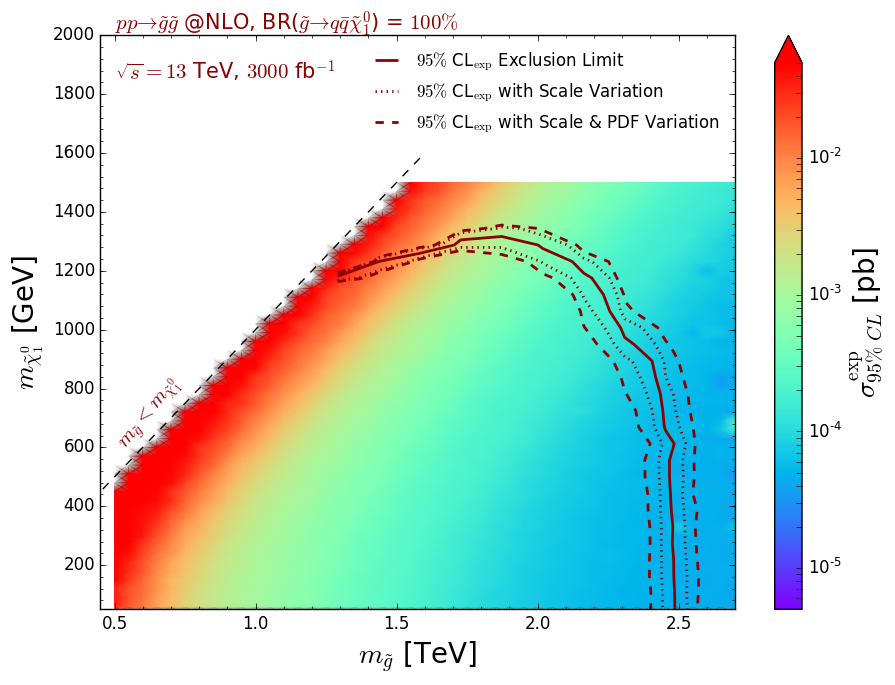}
  \caption{\it Expected constraints on the gluino-neutralino simplified model
    under consideration, represented as 95\% confidence level exclusion contours
    in the $(m_{\tilde g}, m_{\tilde\chi_1^0})$ plane. We present the exclusions
    derived by extrapolating with \madanalysis\ the expectation of the
    ATLAS-SUSY-2016-07 analysis for 36~fb$^{-1}$ of LHC collisions to
    300~fb$^{-1}$ (upper) and 3000~fb$^{-1}$ (lower). In the left panel, we
    extrapolate the uncertainties on the background linearly (\ie~the errors
    are assumed to be dominated by the systematics) while in the right panel, we
    extrapolate them proportionally to the square root of the luminosity
    (\ie~the errors are assumed to be dominated by statistics). The colour
    scheme represents the cross section value excluded at the 95\% confidence
    level for each mass configuration.
  \label{fig:exclusion_HL}}
\end{figure*}

With the goal of assessing the importance of the signal normalisation,
we extract bounds on the model by making use of NNLO$_{\rm approx}$+NNLL rates
(red contour) instead of NLO ones (blue contour), NNLO$_{\rm approx}$+NNLL
predictions being the most precise estimates for gluino-pair production to date.
While still different from what has been used in the ATLAS study, NLO-NLL and
NNLO$_{\rm approx}$+ NNLL predictions are known to be consistent with each
other when theory error bands are accounted
for. This has been documented, in the case of a gluino simplified model in which
all squarks are decoupled, by the LHC Supersymmetry Cross Section Working
Group\footnote{See the webpage
\url{https://twiki.cern.ch/LHCPhysics/SUSYCrossSections13TeVgluglu}.}. We
observe a better agreement with the ATLAS results, showing the important role
played by the new physics signal normalisation in the limit setting procedure.
Large differences of about 5\% on the mass limits are nevertheless still
noticeable, showing that not only the normalisation but also the shape of the
distributions are important ingredients. The ATLAS-SUSY-2016-07 analysis indeed
relies on the $M_{\rm eff}(N)$ variable that is particularly sensitive to the
modelling of the second jet, as $N\geq 2$ for all the analysis signal regions.
In our setup in which NLO matrix elements are matched with parton shower, the
second jet properties are described at the leading-logarithmic accuracy, the
presence of this jet in the event final state solely originating from parton
showering. This contrasts with ATLAS simulations in which LO matrix-element
corrections are included as well, their final merged Monte Carlo signal samples
including the contributions of LO matrix elements for gluino pair-production in
association with two jets. This should motivate the usage of merged NLO
samples matched with parton showers, so that predictions for observables
sensitive to the sub-leading jet activity could be precisely achieved both for
the shapes and the rates. The investigation of the actual impact of such an NLO
multipartonic matrix element merging however goes beyond the scope of this
work.

We also estimate in figure~\ref{fig:exclusion_36}, the impact of the scale and PDF errors on the exclusion
contours. For both \madanalysis\ predictions in which NLO (blue contour) and
more precise NNLO$_{\rm approx}$+NNLL (red
contour) are used for the signal normalisation, we describe the effect of the
scale uncertainties through dotted contours and the one of the combined scale
and parton density uncertainties through dashed contour. It turns out that the
uncertainties on the signal impacts the gluino mass limits by about 50~GeV
 in both cases, the
effect being mostly dominated by scale variations. The reach of the
considered ATLAS-SUSY-2016-07 analysis concerns gluino masses smaller than
a\-bout 1.8~TeV. This corresponds to a mass range where the uncertainty on the
predictions is dominated by the scale variations, as shown in
figure~\ref{fig:gluprod}. The latter indeed shows that the PDF errors (lower
panel of the figure) are at the level of a few percents for $m_{\tilde g} <
1.8$~TeV, the parton density fits being under a very good control for the
corresponding Bjorken-$x$ values.

In order to estimate the reach of this ATLAS supersymmetry search in the context
of the future runs of the LHC, we make use of the framework detailed in
section~\ref{sec:newma5} to extrapolate the results to 300~fb$^{-1}$ and
3000~fb$^{-1}$. As the ATLAS note of ref.~\cite{Aaboud:2017vwy} does not include
detailed and separate information on the systematical and statistical components
of the uncertainties associated with the SM expectation in each signal region,
we consider the two implemented options for their extrapolation to higher
luminosities. More conservative, a linear extrapolation assumes that the error
on the SM background is mostly dominated by its systematical component and
scales proportionally to the luminosity (see the first term in
eq.~\eqref{eq:bgd_xtp}). More aggressive, an extrapolation in which the error
scales proportionally to the
square root of the luminosity (second term of eq.~\eqref{eq:bgd_xtp}) considers
that the background uncertainties are mainly of a statistical origin. The second
option hence naively leads to a more important gain in sensitivity for higher
luminosities, by definition. For all our predictions, we normalise
the signal rates to NLO.

\begin{figure*}
  \centering
  \includegraphics[width=0.48\textwidth]{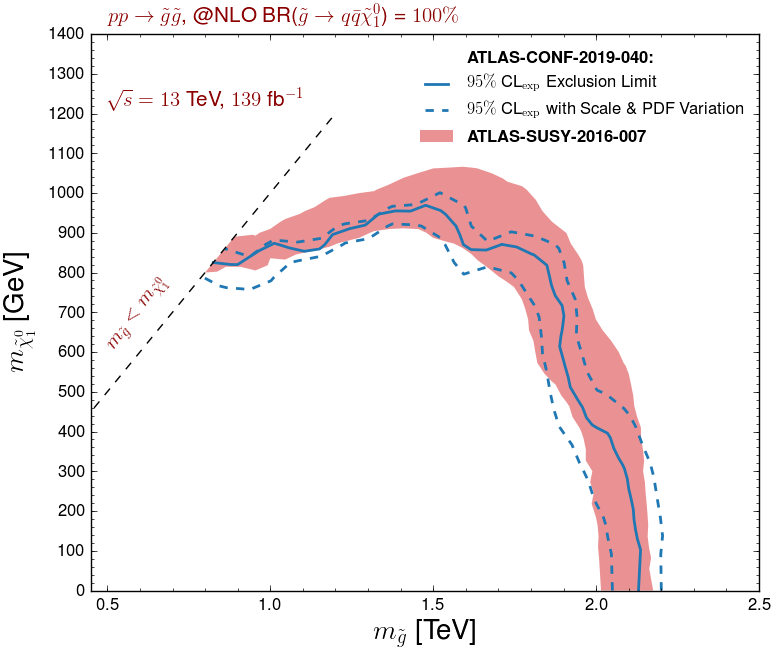}
  \includegraphics[width=0.48\textwidth]{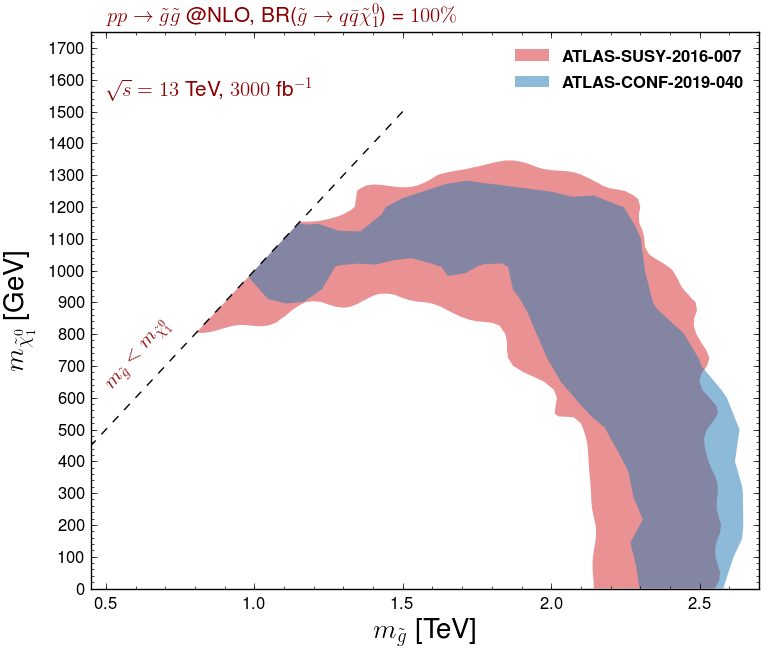}
  \caption{\it Expected constraints on the gluino-neutralino simplified model
    under consideration, represented as 95\% confidence level exclusion contours
    in the $(m_{\tilde g}, m_{\tilde\chi_1^0})$ plane for 139~fb$^{-1}$ (left)
    and 3000~fb$^{-1}$ (right) of proton-proton collisions at a centre-of-mass
    energy of 13~TeV. We compare predictions obtained by recasting the results
    of the ATLAS-CONF-2019-140 analysis (blue lines), which we then extrapolate
    to 3000~fb$^{-1}$ (filled blue area), with those obtained by extrapolating
    the expectation of the ATLAS-SUSY-2016-07 analysis of 36~fb$^{-1}$ of LHC
    data to 139~fb$^{-1}$ and 3000~fb$^{-1}$ (solid red areas). The parameter
    space regions spanned by the various contours correspond to including both
    the PDF and scale uncertainties. The extrapolations are moreover performed
    conservatively (see the text).
  \label{fig:atlas140}}
\end{figure*}

The results are presented in figure~\ref{fig:exclusion_HL},
first, by scaling the background uncertainties linearly to the luminosity (left
panel, assuming that the background errors are dominated by the
systematics), and second, by scaling them proportionally to the square root of
the luminosity (right panel, assuming that the background errors are
dominated by the statistical uncertainties). In all cases, we moreover assess
the impact of the theory errors, the scale and PDF uncertainties being combined
quadratically.

For an extrapolation to 300~fb$^{-1}$ (upper subfigures), the gluino mass limits
are pushed to $2.1-2.2$~TeV for a light bino-like neutralino with
$m_{\tilde\chi_1^0} \lesssim 500$~GeV. The 36~fb$^{-1}$ exclusion is then found
to be improved by about $15-20$\% (or $300-400$~GeV). For such a mass range, the
error on the theoretical predictions is still dominated by the scale variations
(see figure~\ref{fig:gluprod}) and only mildly impacts the exclusion, the effects
reaching a level
of about 5\%. Such a small effect on a mass limit is related to the behaviour of
the cross section with the increasing gluino mass,  that is only
reduced by a factor of a few. Comparing the left and right upper figures, one
can assess the impact of the different treatment for the extrapolation of the
background uncertainties. In the parameter space region under discussion, the
impact is mild, reaching roughly a level of about 5\% on the gluino
mass limit. Such a small effect originates from the small resulting difference
on the background error, that is 3 times smaller in the more aggressive case.
Correspondingly, this allows us to gain a factor of a 3 in cross section, or
equivalently a few hundreds of GeV in terms of a mass reach.

For more compressed scenarios in which the neutralino is heavier
($m_{\tilde\chi_1^0} \gtrsim 800$~GeV) and the gluino lighter ($m_{\tilde g} \in
[1, 1.7]$~TeV), the treatment of the background extrapolation has a quite severe
impact on the bounds on the neutralino mass. A more conservative linear
extrapolation of the background error does not yield any significant change
comparatively to the 36~fb$^{-1}$ case, neutralinos lighter than about 800~GeV
being excluded. However, treating more aggressively the background uncertainties
as being purely statistical,  leads to an important increase in the bounds,
neutralino masses ranging up to about 1~TeV becoming reachable. In those
configurations, the spectra are more compressed and therefore more complicated
to probe than for split configurations, consequently to the fact that the signal
regions are less populated by the supersymmetry signals. A more precisely known
background (with a relatively smaller uncertainty) is therefore crucial for
being able to draw conclusive statements. As found in our results, any
improvement, as little it is, can have a large impact.

In the lower subfigures, we present the results of an extrapolation to
3000~fb$^{-1}$. All above-described effects are emphasised to a larger extent.
The differences in the treatment of the background uncertainties corresponding
to knowing the background more accurately indeed now involve a factor of 10
in precision. A more interesting
aspect concerns the theoretical predictions themselves that turn out to be known
less and less precisely consequently to large parton density uncertainties. The
limits indeed enter a regime in which large Bjorken-$x$ are probed, which
corresponds to PDF uncertainties contributing significantly to the total theory
error. A better knowledge of the parton densities at large $x$ and large scale
is thus mandatory to keep our capacity to probe new physics in this regime.

We have verified that the obtained bounds were compatible with the naive
extrapolations performed by the
\href{http://collider-reach.web.cern.ch/collider-reach}{\sc Collider Reach}%
\footnote{See the webpage
\url{http://collider-reach.web.cern.ch}.} platform that
extracts naive limits of a given collider setup with respect to the reach of a
second collider setup, rescaling the results of the later by ratio of partonic
luminosities. For instance, an 1.8~TeV gluino excluded with 36~fb$^{-1}$ of LHC
collisions would correspond to a $2.4-2.7$~TeV exclusion at 300~fb$^{-1}$. This
is in fair agreement with our findings, after accounting for the fact that {\sc
Collider Reach} uses the NNPDF 2.3 set of parton densities~\cite{Ball:2012cx}, a
set of parton distribution functions whose fit only includes 2010 and 2011 LHC
data, so that important differences are expected, particularly for large
$x$-values.

 Whilst our extrapolations rely on the reinterpretation of an ATLAS
analysis of 36~fb$^{-1}$ of LHC collisions, they are quite robust despite the
small luminosity under consideration. Multijet plus missing transverse energy
studies targeting a monojet-like topology (\ie\ with a hard selection on the
leading jet) are indeed limited by systematics~\cite{Banerjee:2017wxi}, so that
only mild improvements could be expected with a higher luminosity. This is what
has been found in the results of figure~\ref{fig:exclusion_HL}, the bounds being
improved by at most 20\% in mass when going from 300 to 3000~fb$^{-1}$. This
subsequently also implies that the expected sensitivity should be rather
independent of the initially-analysed luminosity. We further demonstrate those
considerations in figure~\ref{fig:atlas140}.

In the left panel of the figure, we extrapolate the results of the
ATLAS-SUSY-2016-07 analysis to the full Run~2 luminosity of 139~fb$^{-1}$, the
theory errors being combined quadratically. In our extrapolation procedure, we
have considered both that the background uncertainties are dominated by the
systematics (linear scaling) and by the statistics (scaling proportional to the
square root of the luminosity). The
two set of results have been merged and presented as the unique envelope of the
exclusion bands derived from the two extrapolation procedures. They could
hence be seen as a conservative theory estimate for the LHC sensitivity at
139~fb$^{-1}$, when estimated from official 36~fb$^{-1}$ results.

The ATLAS-SUSY-2016-07 analysis has been updated last summer as the
ATLAS-CONF-2019-040 analysis~\cite{ATLAS:2019vcq}, so that the most recent and
stringent ATLAS limits on the considered gluino simplified model now encompass
the analysis of the full LHC Run~2 dataset. On the other hand, the updated
analysis has been recently added to the PAD~\cite{atlas_conf_2019_040_ma5}, so
that it can be used within the \madanalysis\ framework for reinterpretation
studies. The corresponding 95\% confidence level contour is shown on the left
panel of figure~\ref{fig:atlas140} (solid blue line), together with the
uncertainty band stemming from combining the scale and PDF uncertainties in
quadrature. In addition, we also present predictions for the bounds as obtained
from an extrapolations of early Run~2 results focusing on 36~fb$^{-1}$ of LHC
data. After accounting for the error bands, the two sets of constraints are
found in good agreement, as expected.

On the right panel of figure~\ref{fig:atlas140}, we consider the two ATLAS
multijet plus missing transverse energy analyses that have been above-mentioned,
namely the early LHC Run~2 ATLAS-SUSY-2016-07 analysis (36~fb$^{-1}$, red) and
the full Run~2 ATLAS-CONF-2019-040 analysis (139~fb$^{-1}$, blue). We
reinterpret their results with \madanalysis, and extrapolate the predictions
that have been obtained for the nominal luminosities of the two analyses to
3000~fb$^{-1}$. The contours shown on the figure are obtained as before, \ie\
by considering independent scalings of the background assuming that it is either
dominated by the systematical or by the statistical uncertainties. The
envelopes of the two exclusion bands (including the theory errors) are then
reported in the figure. The two solid areas presented on the figure are found to
largely overlap and be consistent with each other.

\begin{figure}
  \centering
  \includegraphics[width=0.97\columnwidth]{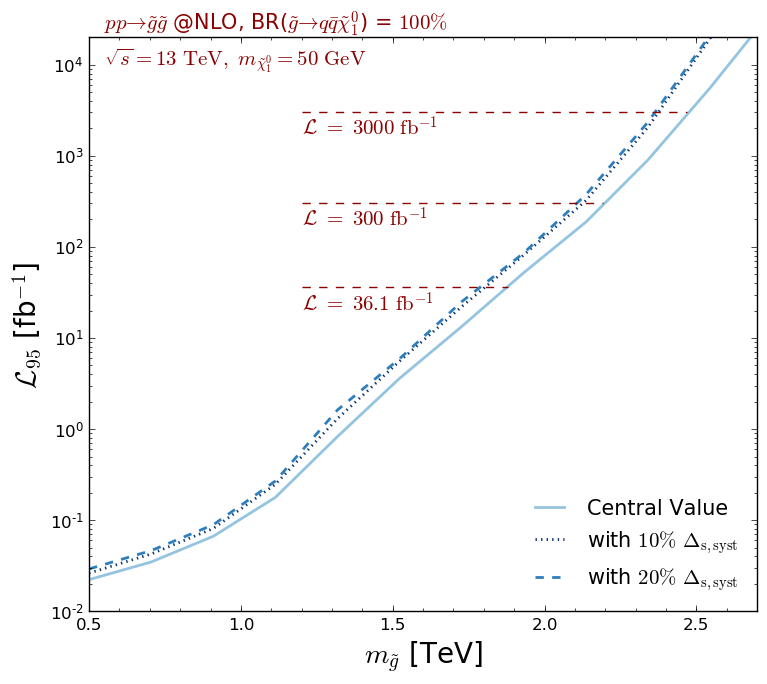}
  \caption{\it Luminosity necessary to exclude, at the 95\% confidence level, a
    given gluino-neutralino new physics setup with the ATLAS-SUSY-16-07
    analysis. We fix the neutralino mass to $m_{\tilde\chi_1^0}=50$~GeV, assume
    that the uncertainties on the background are dominated by their statistical
    component, and include systematical uncertainties on the signal of 0\% (solid
    line), 10\% (dotted line) and 20\% (dashed line).
  \label{fig:lumi95}}
\end{figure}

In figure~\ref{fig:lumi95}, we make use of the \madanalysis\ infrastructure to
estimate, for various benchmark points, the luminosity ${\cal L}_{95}$ that is
required to exclude the scenario at the 95\% confidence level. We still consider
the ATLAS-SUS-2016-07 analysis, fix the neutralino mass to 50~GeV and let the
gluino mass vary. We compute ${\cal L}_{95}$ for two choices of systematics on
the signal (combined in both cases with the theory errors quadratically), namely
10\% (dotted
line) and 20\% (dashed line), and compare the predictions with the central value
where the signal is perfectly known (solid line). In those calculations, we
scale the error on the background proportionally to the square root of the
luminosity, as if it was mainly dominated by its statistical component. Our
analysis first shows that light gluinos with masses smaller than about 1.5~TeV
can be excluded with a luminosity ${\cal L}_{95}$ of a few fb$^{-1}$, as
confirmed by the early Run~2 ATLAS search of ref.~\cite{Aaboud:2016zdn} that
consists of the 3.2~fb$^{-1}$ version of the ATLAS-SUSY-2016-07 analysis.
The steep fall of the cross section with an increasing gluino mass moreover
implies that the high-luminosity LHC reach of the analysis under consideration
will be limited to gluinos of about 2.5~TeV, a bound that could be reduced by
about 10\% if the systematics on the signal are of about $10-20$\%. This
order of magnitude has been found to agree with older ATLAS
estimates~\cite{ATLAS:2019vcq}.

\section{Squark and neutralino mass limits}\label{sec:squarks}
\begin{figure}
 \centering
 \includegraphics[width=0.70\columnwidth]{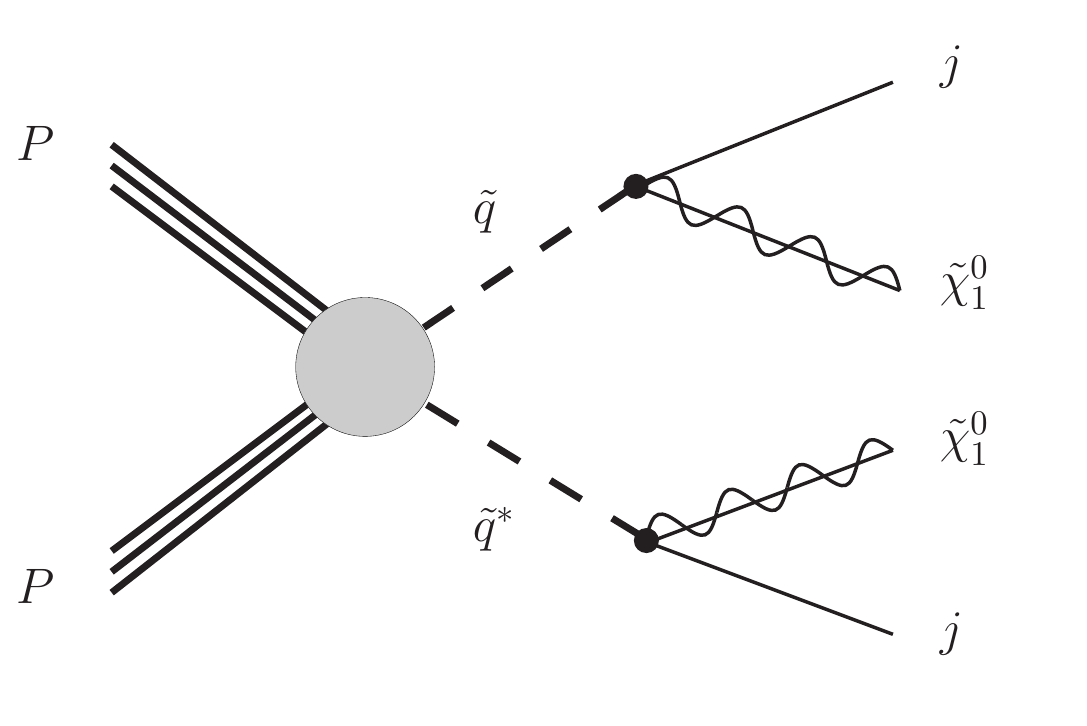}
 \caption{\it Generic Feynman diagram associated with the production
   and decay of a pair of squarks in the considered MSSM-inspired squark
   simplified model. The figure has been produced with the help of the
   {\sc JaxoDraw} package~\cite{Binosi:2008ig}.
 \label{fig:squarkdiag}}
\end{figure}

In this section, we consider a second class of simplified models inspired by the
MSSM that is widely studied in the context of the LHC searches for new physics.
As in section~\ref{sec:GluProd}, all superpartners, except for two under investigation,  are decoupled.  This time, these are taken to be a squark and the
lightest neutralino. In practice, we hence supplement the SM field content by
one species of first generation squark $\tilde q$ and the lightest neutralino
$\tilde\chi_1^0$,  assumed to be bino-like. In this configuration, squarks can
be pair-produced through standard QCD interactions, and then each decays into the
lightest neutralino and an up quark, as illustrated by the generic Feynman
diagram of figure~\ref{fig:squarkdiag}. Such a parton-level final state
comprised of two quarks and two invisible neutralinos therefore manifests
itself, after parton showering and hadronisation, as a multijet plus missing
transverse energy topology.

The ATLAS analyses considered in section~\ref{sec:GluProd}, targeting multijet
plus missing energy signs of new physics, are therefore appropriate to put
constraints on the model under consideration. Those analyses indeed include not
only signal regions dedicated to probe final state featuring a large jet
multiplicity (that are thus ideal to target the previously considered gluino
simplified model), but also include signal regions targeting signals exhibiting
a smaller jet multiplicity (that are thus excellent probes for the present
squark simplified model). In the following, we only make use of the most recent
search, ATLAS-CONF-2019-140~\cite{ATLAS:2019vcq}.

\begin{figure}
  \centering
  \includegraphics[width=0.97\columnwidth]{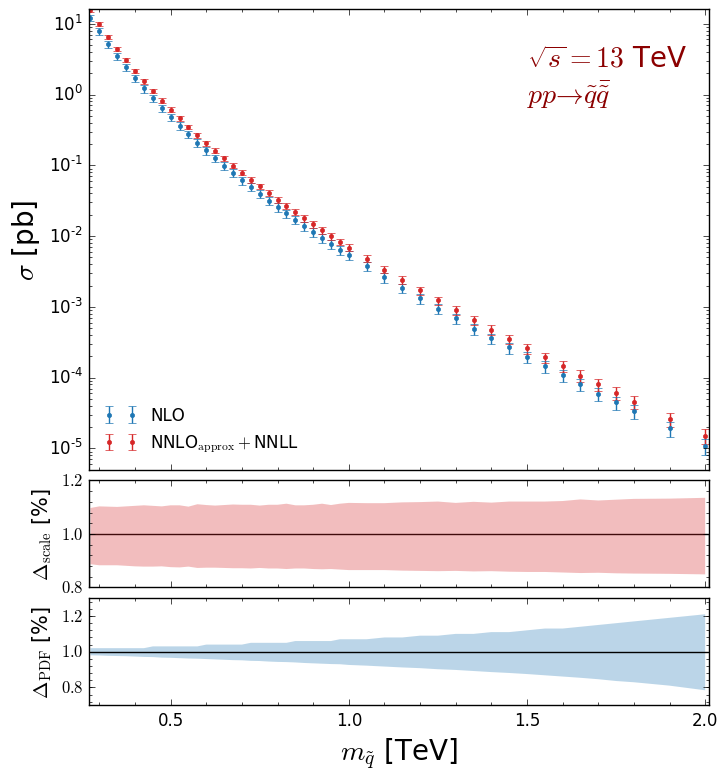}
  \caption{\it Total NLO (blue) and approximate NNLO+NNLL (red)
    cross section (upper panel) for squark pair production in proton-proton
    collisions at a centre-of-mass energy of 13~TeV. The error bars represent
    the quadratic sum of the scale and PDF uncertainties. In the middle and
    lower panels of the figure, we report the NLO scale and PDF uncertainties
    respectively, after normalising the results to the central NLO cross section
    value.
  \label{fig:sq_xsec}}
\end{figure}

\begin{figure*}
  \centering
  \includegraphics[width=0.48\textwidth]{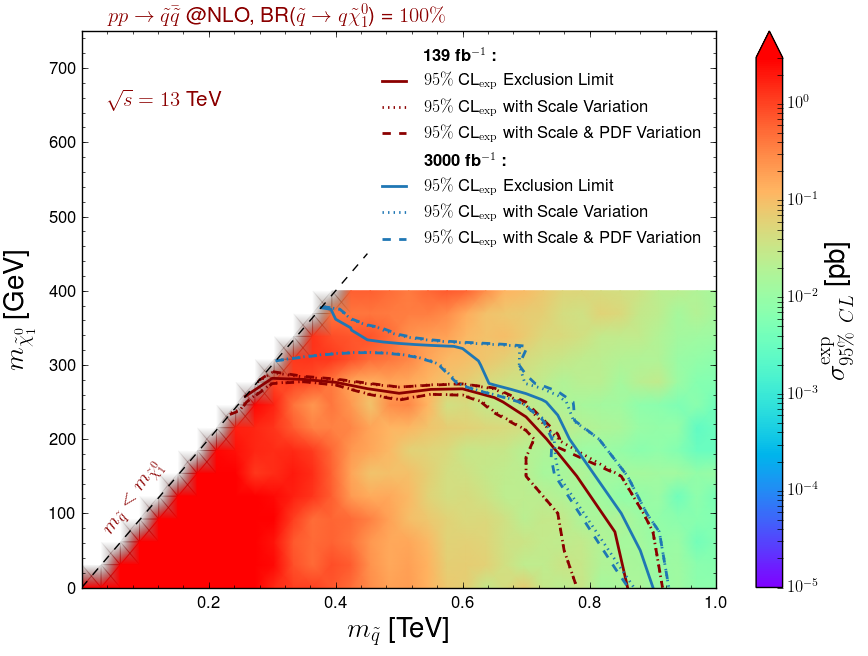}
  \includegraphics[width=0.48\textwidth]{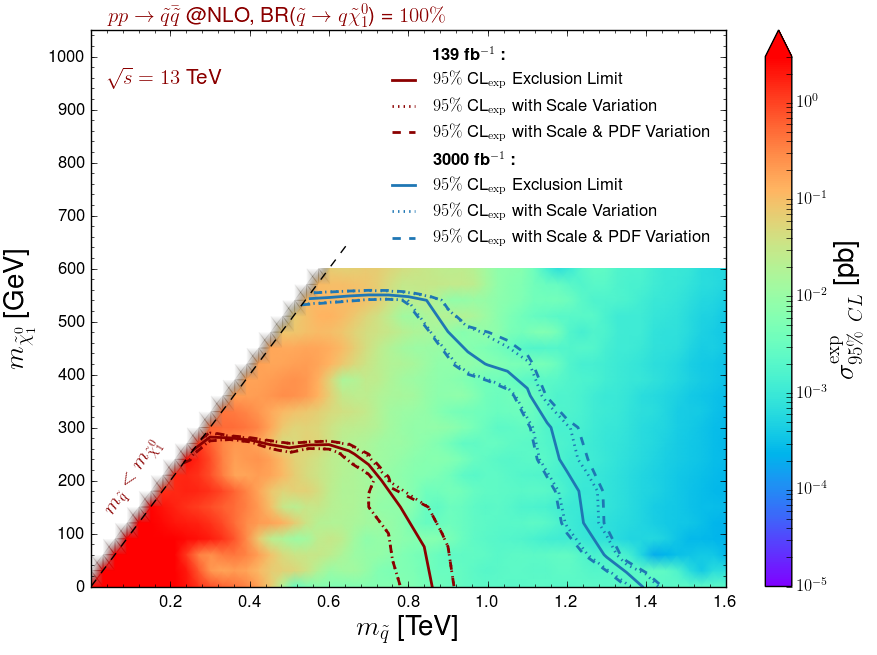}
  \caption{\it Expected constraints on the squark-neutralino simplified model
    under consideration, represented as 95\% confidence level exclusion contours
    in the $(m_{\tilde q}, m_{\tilde\chi_1^0})$ plane for 139~fb$^{-1}$ (red)
    and 3000~fb$^{-1}$ (blue) of proton-proton collisions at a centre-of-mass
    energy of 13~TeV. We derive those bounds with the ATLAS-CONF-2019-140
    implementation in \madanalysis\ and extrapolate the uncertainties on the
    background as if they are systematically-dominated (left, scaling
    proportional to the luminosity ) or statistically-dominated (right, scaling
    proportional to the sqare root of the luminosity).
\label{fig:SqProd_139_HL}}
\end{figure*}

Making use of the same simulation setup as in section~\ref{sec:GluProd}, we
study the LHC sensitivity to this model after the full Run~2 and present the
expectation of its high-luminosity operation run. Our results are derived from
simulations at the NLO+PS accuracy, so that our signal samples are normalised at
the NLO accuracy. The rates that we employ in the following are depicted in the
upper panel of figure~\ref{fig:sq_xsec}, where we show total NLO-accurate
squark-pair production cross sections as returned by \madgraph\ when using
the MSSM implementation developed in ref.~\cite{Frixione:2019fxg} and the NLO
set of NNPDF~3.0 parton densities~\cite{Ball:2014uwa} (blue). Predictions are
given for squark masses ranging from 250~GeV to 2~TeV and include theory errors
that we estimate by adding scale and PDF uncertainties in quadrature. Those
uncertainties are further described more precisely in the middle and lower
panels of the figure, where they are given after normalising the results to
the central NLO cross section value for each mass point.

We obtain cross sections that vary from 10~pb for $m_{\tilde q}\sim 250$~GeV
to 0.01~fb for 2~TeV squarks. They are two orders of magnitude lower than
in the gluino case for a specific mass value, as expected from the fact that squarks are scalars
 and are colour triplets and not octets. Scale
uncertainties are found to be independent of the squark mass for the considered
$m_{\tilde q}$ range, and are of about 15\% (middle panel of the figure). In
contrast, the PDF errors strongly depend on the squark mass $m_{\tilde q}$
(lower panel of the figure), as they are correlated with the associated
Bjorken-$x$ regime. They are of a few percents and thus subleading for small
$m_{\tilde q}$ values, and grow for increasing squark masses, eventually
reaching 20\% for $m_{\tilde q} = 2$~TeV. For larger and larger $x$-values (and
thus larger and larger $m_{\tilde q}$), the quark-antiquark contributions to the
cross section play a bigger and bigger role. Simultaneously, the impact of the
Bjorken-$x$ regime in which the PDF sets are more poorly constrained by data
gets more important.

As in section~\ref{sec:GluProd}, we compare our predictions
to the cross section values usually employed by the LHC collaborations (red
curve), as reported by the LHC Supersymmetry Cross Section Working
Group~\cite{Borschensky:2014cia}. The latter are however only provided for a
simplified model in which all squarks except the two stop squarks are
mass-degenerate. We therefore normalise the NNLO$_{\rm approx}$+NNLL results
by an extra factor of 1/10, which should be a fair enough approximation for
small squark masses. Nevertheless, as both NLO and NNLO$_{\rm approx}$+NNLL
predictions are consistent, we consider (exact) NLO rates in the following.

In figure~\ref{fig:SqProd_139_HL}, we reinterpret the results of the
ATLAS-CONF-2019-040 analysis with \madanalysis\ and pre\-sent the expected
exclusion contours both at the nominal luminosity of 139~fb$^{-1}$,  after
extrapolating the findings to 3000~fb$^{-1}$, using for each point the region
yielding the best expected sensitivity. Neutralino masses below about 300~GeV
are currently (\ie\ for a luminosity of 139~fb$^{-1}$) excluded, for squark
masses ranging up to about 900~GeV. This may seem to contrast by a factor of
about 2 with the current bounds on this class of simplified model set by the
ATLAS collaboration~\cite{ATLAS:2019vcq}. This is however not surprising as the
collaboration only interprets its results for a simplified model in which the
superpartner spectrum exhibits 10 mass-denegerate left-handed and right-handed
squarks (\ie\ all squarks except the two stop squarks are degenerate). The
corresponding signal cross sections are therefore about 10 times larger, so that
much stronger limits could be extracted. In comparison with final Run~2 CMS
results~\cite{Sirunyan:2019xwh,Sirunyan:2019ctn} for which result
interpretations both for eight mass-degenerate squarks and a single squark
species are provided, we obtain more conservative bounds that are roughly 20\%
weaker in terms of excluded masses. When accounting for the uncertainty bands,
our predictions agree with the experimental findings, as the uncertainty bands
overlap.

Extrapolating the results to a luminosity of 3000~fb$^{-1}$, \ie\ expected
luminosity of the high-luminosity phase of the LHC, we obtain expected bounds which are 
improved quite a bit. The magnitude of the improvement is found strongly related
to how the background uncertainties will be controlled, as visible by comparing
the cur\-ves corresponding to 3000~fb$^{-1}$ (blue) in the two panels of the
figure. Assuming that the background is dominated by the systematics or the
statistics change the results by more than 40\%.

\section{Sensitivity to simplified $s$-channel dark matter models}
\label{sec:dm}

In this section, we investigate the sensitivity of the LHC to a simplified dark
matter (DM) model. We assume that DM is described by a massive Dirac fermionic
particle $X$ that communicates with the Standard Model through the exchange of a
spin-1 mediator $Y$. Motivated by models with an extended gauge group, we
consider that the mediator couples only either to a pair of DM particles, or to
a pair of SM fermions. Such a configuration is typical from the so-called
$s$-channel dark matter models~\cite{Backovic:2015soa,Abercrombie:2015wmb}. In
this class of scenarios, DM can only be pair-produced at colliders, from the
scattering of a pair of SM quarks and through the $s$-channel exchange of the
mediator.

The corresponding Lagrangian can generically be written as
\be\bsp
  {\cal L}= &\ {\cal L}_{\rm SM} + {\cal L}_{\rm kin} +
     \bar{X} \gamma_\mu \big[g^V_X + g^A_X \gamma_5 \big] X\ Y^\mu \\ &\ \ +
     \sum_q \Big\{\bar{q} \gamma_\mu \big[ g^V_q + g^A_q \gamma_5 \big] q \Big\}
     Y^\mu
\esp\ee
where ${\cal L}_{\rm SM}$ refers to the SM Lagrangian and ${\cal L}_{\rm kin}$
contains gauge-invariant kinetic and mass terms for all new fields. The next
term includes the vector and axial-vector interactions of the mediator with DM,
their strength being denoted by $g^V_X$ and $g^A_X$ respectively, and the last
term focus on the mediator interactions with the SM quarks. The latter are
assumed universal and flavour-independent, their strength being $g^V_q$ and
$g^A_q$ in the vector and axial-vector case respectively, regardless of the
quark flavour.

In our analysis, we focus on two further simplified scenarios originating from
that model. In a first case (that we label {\bf S1}), the mediator couplings are
taken as of a vectorial nature, whilst in the second case (that we label
{\bf S2}), they are taken as of an axial-vectorial nature. In other words, the
two scenarios are defined as
\be
  \text{\bf S1}:\ g_q^A = g_X^A = 0\ ; \qquad
  \text{\bf S2}:\ g_q^V = g_X^V = 0 \ .\label{eq:scenario}
\ee

In order to study the sensitivity of the LHC to these two classes of scenarios,
we make use of the publicly available\footnote{See the webpage
\url{http://feynrules.irmp.ucl.ac.be/wiki/DMsimp}.} implementation of the model
in the \fr\ package~\cite{Alloul:2013bka} introduced in
ref.~\cite{Backovic:2015soa}, as well as of the corresponding public
UFO~\cite{Degrande:2011ua} library. As in the previous sections, hard scattering
events are generated at the NLO accuracy in QCD with
\madgraph~\cite{Alwall:2014hca} and then matched with parton showering and
hadronisation as performed by \pythia~\cite{Sjostrand:2014zea}. In our
simulations, the matrix elements are convoluted with the NLO set of NNPDF~3.0
parton densities~\cite{Ball:2014uwa}. We derive the LHC sensitivity to the model
by considering the associated production of a pair of dark matter particles with
jets, a signature targeted by the ATLAS-CONF-2019-040
analysis~\cite{ATLAS:2019vcq} introduced in the previous sections. This ATLAS
study searches for new phenomena in a luminosity of 139~fb$^{-1}$ of LHC data at
a centre-of-mass energy of 13~TeV, investigating events featuring at least two
hard jets and a potential subleading jet activity.

\begin{figure}
 \centering
 \includegraphics[width=0.70\columnwidth]{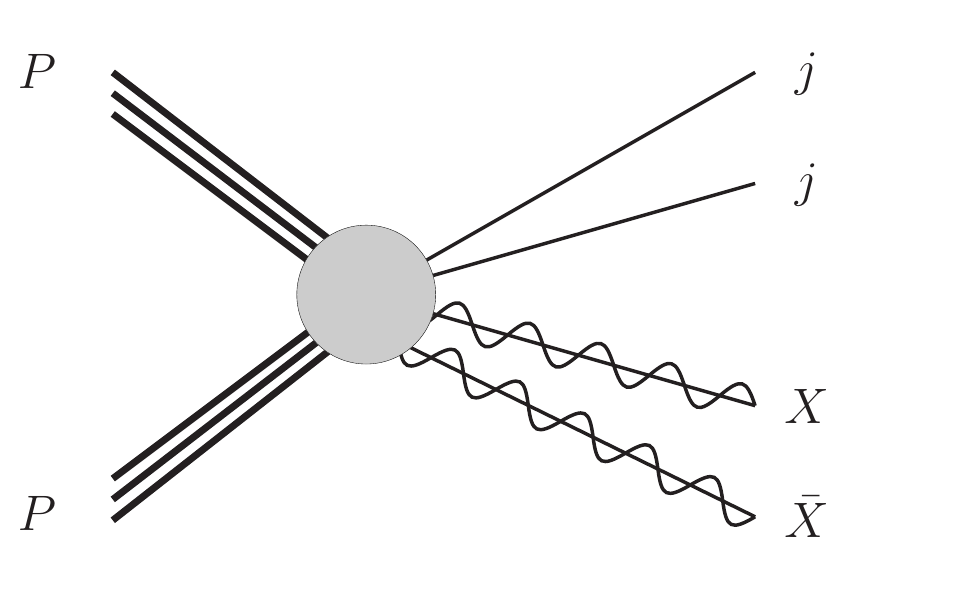}
 \caption{\it Generic Feynman diagram associated with the production of a pair
   of dark matter particles $X$ in association with two hard jets. The figure
   has been produced with the help of the {\sc JaxoDraw}
   package~\cite{Binosi:2008ig}.\label{fig:DMsimpdiag}}
\end{figure}

As the analysis selection requires at least two very hard jets, we consider as a
hard-scattering process the production of a pair of DM particles with two hard
jets, as sketched in figure~\ref{fig:DMsimpdiag}. Moreover, we impose two
conservative (with respect to the ATLAS analysis) generator-level selections. We
constrain the transverse momentum of the hardest of the jets to satisfy
$p_T>150$~GeV, and the parton-level missing transverse energy (\ie\ the
transverse energy of the vector sum of the transverse momenta of the two DM
particles) to fulfil $\slashed{E}_T >150$~GeV. Moreover, the reference
renormalisation and factorisation scales are set to the mass of the mediator
$m_Y$, and we estimate the associated uncertainties as usual, by independently
varying the two scales by a factor of 2 up and down around the central scale
choice.

We begin with considering a series of scenarios featuring light dark matter,
\ie\ with a dark matter mass $m_X$ fixed to 100~GeV. The mediator mass is
kept free to vary in the [0.3, 2]~TeV range. In table~\ref{tab:myscan},
we present the sensitivity of the ATLAS-CONF-2019-040
analysis to those scenarios, both at the nominal luminosity of 139~fb$^{-1}$
and for the high-luminosity LHC run (with 3000~fb$^{-1}$). For each spectrum
configuration, we show NLO signal cross sections (second and fifth columns for
the {\bf S1} and {\bf S2} benchmarks respectively), as obtained following the
simulation setup described above and for couplings obeying to
eq.~\eqref{eq:scenario}. Moreover, those predictions are obtained after fixing
the remaining non-vanishing free parameters to the reference values
\be\bsp
  \text{\bf S1}:&\ g_q^V = 0.25\ ,\ g_X^V = 1\ ; \\
  \text{\bf S2}:&\ g_q^A = 0.25\ ,\ g_X^A = 1 \ ,
\esp\label{eq:scenario2}\ee
which consist in one of the benchmarks studied by the LHC dark matter working
group~\cite{Abercrombie:2015wmb}.

\begin{table*}
  \renewcommand{\arraystretch}{1.6}
  \setlength\tabcolsep{4pt}
  \begin{center}
  \begin{tabular}{c|ccc|ccc}
    & \multicolumn{3}{c|}{Vector couplings ({\bf S1})} &
      \multicolumn{3}{c} {Axial-vector couplings ({\bf S2})}\\
        $m_{Y}$ [TeV] &
        $\sigma_{\rm NLO}$ [pb] & $\sigma_{95} $ [pb]  & $|g_{95}| $ &
        $\sigma_{\rm NLO}$ [pb]& $\sigma_{95}$ [pb] & $|g_{95}|$ \\ \hline
      $0.3$ &$ 19.45^{+35.4\%\ +1.2\%}_{-77.1\%\ -1.2\%} $ & $1.553 (1.318)$ & $ 0.532^{+0.06}_{-0.07} \left(0.510^{+0.06}_{-0.07}\right) $& $ 9.21^{+30.9\%\ +1.2\%}_{-68.4\%\ -1.2\%} $ & $1.015 (0.700)$ & $ 0.576^{+0.06}_{-0.07} \left(0.525^{+0.05}_{-0.06}\right) $\\
      $0.5$ &$ 12.19^{+15.5\%\ +1.1\%}_{-39.3\%\ -1.1\%} $ & $0.667 (0.568)$ & $ 0.484^{+0.02}_{-0.04} \left(0.465^{+0.02}_{-0.04}\right) $& $ 9.73^{+15.8\%\ +1.3\%}_{-39.8\%\ -1.3\%} $ & $0.643 (0.545)$ & $ 0.507^{+0.02}_{-0.04} \left(0.486^{+0.02}_{-0.04}\right) $\\ 
      $0.7$ &$ 7.05^{+10.3\%\ +1.2\%}_{-29.5\%\ -1.2\%} $ & $0.368 (0.311)$ & $ 0.478^{+0.01}_{-0.03} \left(0.458^{+0.01}_{-0.03}\right) $& $ 6.41^{+9.2\%\ +1.3\%}_{-27.3\%\ -1.3\%} $ & $0.333 (0.285)$ & $ 0.477^{+0.01}_{-0.03} \left(0.459^{+0.01}_{-0.03}\right) $\\ 
      $0.8$ &$ 5.37^{+8.3\%\ +1.4\%}_{-25.8\%\ -1.4\%} $ & $0.312 (0.266)$ & $ 0.491^{+0.01}_{-0.03} \left(0.472^{+0.01}_{-0.03}\right) $& $ 5.22^{+6.3\%\ +1.2\%}_{-22.0\%\ -1.2\%} $ & $0.278 (0.234)$ & $ 0.480^{+0.01}_{-0.02} \left(0.460^{+0.01}_{-0.02}\right) $\\ 
      $0.9$ &$ 4.15^{+6.1\%\ +1.3\%}_{-21.7\%\ -1.3\%} $ & $0.242 (0.169)$ & $ 0.491^{+0.01}_{-0.02} \left(0.449^{+0.01}_{-0.02}\right) $& $ 4.13^{+5.5\%\ +1.3\%}_{-18.9\%\ -1.3\%} $ & $0.241 (0.205)$ & $ 0.491^{+0.01}_{-0.02} \left(0.472^{+0.01}_{-0.02}\right) $\\ 
      $1.0$ &$ 3.30^{+5.2\%\ +1.6\%}_{-20.0\%\ -1.6\%} $ & $0.224 (0.189)$ & $ 0.511^{+0.01}_{-0.02} \left(0.490^{+0.01}_{-0.02}\right) $& $ 3.39^{+4.7\%\ +1.6\%}_{-17.0\%\ -1.6\%} $ & $0.221 (0.188)$ & $ 0.505^{+0.01}_{-0.02} \left(0.485^{+0.01}_{-0.02}\right) $\\ 
      $1.2$ &$ 2.16^{+4.0\%\ +1.7\%}_{-16.6\%\ -1.7\%} $ & $0.204 (0.174)$ & $ 0.554^{+0.01}_{-0.02} \left(0.533^{+0.01}_{-0.02}\right) $& $ 2.17^{+3.9\%\ +1.7\%}_{-15.0\%\ -1.7\%} $ & $0.175 (0.148)$ & $ 0.533^{+0.01}_{-0.02} \left(0.511^{+0.01}_{-0.02}\right) $\\ 
      $1.4$ &$ 1.44^{+3.7\%\ +2.3\%}_{-13.5\%\ -2.3\%} $ & $0.139 (0.118)$ & $ 0.557^{+0.01}_{-0.02} \left(0.535^{+0.01}_{-0.02}\right) $& $ 1.42^{+2.5\%\ +1.9\%}_{-11.1\%\ -1.9\%} $ & $0.142 (0.120)$ & $ 0.562^{+0.00}_{-0.01} \left(0.539^{+0.00}_{-0.01}\right) $\\ 
      $1.5$ &$ 1.15^{+2.9\%\ +2.1\%}_{-11.9\%\ -2.1\%} $ & $0.139 (0.117)$ & $ 0.590^{+0.01}_{-0.02} \left(0.566^{+0.01}_{-0.02}\right) $& $ 1.15^{+2.6\%\ +2.3\%}_{-11.0\%\ -2.3\%} $ & $0.127 (0.108)$ & $ 0.576^{+0.01}_{-0.02} \left(0.554^{+0.00}_{-0.01}\right) $\\ 
      $1.8$ &$ 0.63^{+2.1\%\ +2.5\%}_{-8.5\%\ -2.5\%} $ & $0.121 (0.103)$ & $ 0.662^{+0.01}_{-0.01} \left(0.636^{+0.01}_{-0.01}\right) $& $ 0.66^{+1.9\%\ +2.6\%}_{-7.8\%\ -2.6\%} $ & $0.133 (0.112)$ & $ 0.672^{+0.01}_{-0.01} \left(0.643^{+0.01}_{-0.01}\right) $\\ 
      $2.0$ &$ 0.44^{+2.1\%\ +2.9\%}_{-8.6\%\ -2.9\%} $ & $0.104 (0.089)$ & $ 0.699^{+0.01}_{-0.02} \left(0.671^{+0.01}_{-0.01}\right) $& $ 0.44^{+1.6\%\ +3.1\%}_{-6.4\%\ -3.1\%} $ & $0.095 (0.081)$ & $ 0.680^{+0.01}_{-0.01} \left(0.653^{+0.01}_{-0.01}\right) $
  \end{tabular}
  \caption{\it
    Expected constraints on various light dark matter $s$-channel
    scenarios. The dark matter mass is fixed to $m_X=100$~GeV and the couplings
    satisfy eq.~\eqref{eq:scenario}. Reference NLO cross sections (second and
    fifth columns) are provided for a case where the remaining free couplings
    are set as in eq.~\eqref{eq:scenario2}, and can be compared with the
    95\% confidence level limits expected from the reinterpretation of the
    ATLAS-CONF-2019-040 analysis of (third and sixth columns). Our results
    are present both at the nominal luminosity of 139~fb$^{-1}$ and after being
    extrapolated to 3000~fb$^{-1}$ assuming systematics-dominated uncertainties
    (in parentheses). Those bounds are
    also translated into a bound on the couplings for a $g_q=g_X$ configuration
    (fourth and seventh columns).
  \label{tab:myscan}}
  \end{center}
\end{table*}

We first assess the LHC sensitivity to each point for the two considered
luminosities in terms of the signal cross section that is reachable at the LHC
$\sigma_{95}$ (third and sixth columns of table~\ref{tab:myscan} for the
{\bf S1} and {\bf S2} benchmarks respectively) by reinterpreting, with
\madanalysis, the results of the ATLAS-CONF-2019-040 analysis. Second, we
translate the cross section limits that we have obtained into a bound on a
universal new physics coupling strength $g_{95}$ that is defined for scenarios
in which
\be
  g_q=g_X\ .
\label{eq:scenario3}\ee
Moreover, we provide the $g_{95}$ limits together with the theory uncertainty
stemming from scale and PDF variations (fourth and seventh column of the table).
The most stringent bounds on the model originate from a single signal region of
the analysis in which, the effective mass $M_{\rm eff}$ is imposed to be larger
than 2.2~TeV. Such a cut is applied together with looser cuts on the jet
properties, as compared with other signal regions featuring smaller effective
masses.

For fixed vector couplings ({\bf S1} scenarios), the NLO cross section
$\sigma_{\rm NLO}$ decreases when the mediator mass increases and spans a range
extending from about 450~fb for heavy mediators with a mass of about 2~TeV, to
more than 10~pb for mediators lighter than 500~GeV. Those values and this
steeply-falling behaviour are mainly driven by the heavy mass of the
mediator as compared with the small dark matter mass. Larger cross sections are
indeed obtained for smaller mediator masses as we lie closer to the resonant
regime in which $m_Y \sim 2 m_X$. The cross section that is expected to be
excluded at the 95\% confidence level also falls down with $m_Y$, although the
slope is much flatter. Moreover, $\sigma_{95} < \sigma_{\rm NLO}$. Consequently,
all scenarios defined by the coupling assumptions of eq.~\eqref{eq:scenario} and
eq.~\eqref{eq:scenario2} are excluded, already with the present full Run~2
luminosity.

Relaxing the coupling definitions of eq.~\eqref{eq:scenario2} and replacing it
by the universal coupling constraint of eq.~\eqref{eq:scenario3}, it turns out
that couplings of 0.4--0.7 are excluded over the entire mass range, the best
limits being obtained for scenarios featuring sub-TeV mediators and a spectrum
such that one lies far enough from the resonant regime. In the latter case, the
analysis is less sensitive as a consequence of the associated softer final state
objects populating the signal events. The overall weak dependence of the
excluded coupling on the mediator mass stems from various interplaying effects.
First, the cross section has a quartic dependence on the couplings, so that
a small coupling change leads to a large modification of the cross section.
Second, there is a strong interplay between the mediator mass and the dark
matter mass (\ie\ if ones lies for enough from the resonant regime) and the
kinematical configuration probed by the analysis cuts, especially for light
mediators.

In the heavy-mediator regime, considering {\bf S2} scenarios featuring
axial-vector mediator couplings leads to very similar results. In this limit,
the relevant matrix elements are insensitive to the mediator nature. On the
contrary, when one approaches the resonant regime, significant chan\-ges arise:
The cross section turns out to be suppressed relatively to the vector {\bf S1}
scenario. This originates from the impact of the threshold regime that plays a
larger and larger role for smaller and smaller masses. At threshold, the pair of
dark matter particles is organised into a $^3P_1$ state, and not into a $^3S_1$
configuration as in the {\bf S1} scenario. Consequently, signal cross sections
are relatively suppressed. The small increase in cross section for low $m_Y$
values in the {\bf S2} case hence stems from those threshold effects that are
more and more tamed when one gets further from threshold, as well as from the
cut on the leading jet of 150~GeV. As in the {\bf S1}
scenario, the entire mass range is excluded by the ATLAS-CONF-2019-040 analysis,
which translates in the exclusion of couplings in the 0.4--0.7 ballpark for the
considered mediator mass range.

Finally, those bounds are expected to only be sightly improved, by about 4--9\%,
after including 3000~fb$^{-1}$ of data for both scenarios. This is related to
the systematical dominance of the uncertainties on the background, as we have
chosen to scale it under that assumption, so that more
luminosity will not bring much compared with the Run~2 results. Moreover, we
observe that the results are plagued by quite modest theoretical uncertainties
at the $g_{95}$ level (by virtue of the quartic dependence of the matrix element
of the coupling).

\begin{table*}
  \begin{center}
  \renewcommand{\arraystretch}{1.6}
  \setlength\tabcolsep{2pt}
  \begin{tabular}{c|c c c|c c c}
    & \multicolumn{3}{c|}{ Vector coupling ({\bf S1})} &
      \multicolumn{3}{c}{ Axial-vector coupling ({\bf S2})}\\
      $m_{X}$ [GeV] & $\sigma_{\rm NLO}$ [pb] & $\sigma_{95} $ [pb]  & $|g_{95}|$
      & $\sigma_{\rm NLO}$ [pb] & $\sigma_{95}$ [pb]& $|g_{95}|$ \\ \hline
      $200$ &$ 1.19^{+3.0\%\ +2.1\%}_{-12.2\%\ -2.1\%} $ & $0.149 (0.126)$ & $ 0.595^{+0.006}_{-0.02} \left(0.571^{+0.005}_{-0.02}\right) $& $ 1.11^{+2.5\%\ +2.2\%}_{-10.7\%\ -2.2\%} $ & $0.148 (0.125)$ & $ 0.605^{+0.005}_{-0.02} \left(0.580^{+0.005}_{-0.01}\right) $\\ 
      $350$ &$ 1.17^{+3.0\%\ +2.2\%}_{-12.5\%\ -2.2\%} $ & $0.115 (0.098)$ & $ 0.560^{+0.005}_{-0.02} \left(0.538^{+0.005}_{-0.02}\right) $& $ 0.85^{+2.5\%\ +2.1\%}_{-10.1\%\ -2.1\%} $ & $0.129 (0.109)$ & $ 0.624^{+0.005}_{-0.02} \left(0.598^{+0.005}_{-0.01}\right) $\\ 
      $500$ &$ 1.10^{+3.5\%\ +2.2\%}_{-12.6\%\ -2.2\%} $ & $0.143 (0.122)$ & $ 0.601^{+0.006}_{-0.02} \left(0.577^{+0.006}_{-0.02}\right) $& $ 0.51^{+2.7\%\ +2.2\%}_{-10.6\%\ -2.2\%} $ & $0.135 (0.114)$ & $ 0.715^{+0.006}_{-0.02} \left(0.685^{+0.006}_{-0.02}\right) $\\ 
      $650$ &$ 0.82^{+3.2\%\ +2.1\%}_{-13.2\%\ -2.1\%} $ & $0.149 (0.127)$ & $ 0.653^{+0.006}_{-0.02} \left(0.627^{+0.006}_{-0.02}\right) $& $ 0.15^{+2.7\%\ +2.4\%}_{-9.7\%\ -2.4\%} $ & $0.143 (0.121)$ & $ 0.982^{+0.009}_{-0.02} \left(0.941^{+0.009}_{-0.02}\right) $\\ 
      $800$ &$ 0.006^{+3.3\%\ +2.8\%}_{-13.8\%\ -2.8\%} $ & $0.131 (0.110)$ & $ 2.171^{+0.02}_{-0.07} \left(2.075^{+0.02}_{-0.07}\right) $& $ 0.0009^{+2.3\%\ +3.2\%}_{-11.4\%\ -3.2\%} $ & $0.121 (0.104)$ & $ 3.456^{+0.03}_{-0.10} \left(3.322^{+0.03}_{-0.09}\right) $\\ 
      $900$ &$ 0.001^{+3.5\%\ +3.5\%}_{-14.7\%\ -3.5\%} $ & $0.107 (0.091)$ & $ 2.986^{+0.04}_{-0.10} \left(2.863^{+0.04}_{-0.10}\right) $& $ 0.0002^{+3.1\%\ +3.5\%}_{-13.4\%\ -3.5\%} $ & $0.110 (0.093)$ & $ 4.600^{+0.06}_{-0.15} \left(4.412^{+0.05}_{-0.14}\right) $\\ 
  \end{tabular}
  \caption{\it
    Same as table~\ref{tab:myscan}, but for a scenario in which $m_X$
    is free and $m_Y$ has been set to 1.5~TeV.\label{tab:mxscan}}
  \end{center}
\end{table*}

In table~\ref{tab:mxscan}, we consider a new class of scenarios. This time, the
mediator mass $m_Y$ is fixed to 1.5~TeV and we vary the dark matter mass $m_X$
from 200 to 900 GeV.

We first consider scenarios with couplings satisfying eqs.~\eqref{eq:scenario}
and \eqref{eq:scenario2}. We evaluate fiducial NLO cross sections for the
different considered mass spectra (second and fifth columns of the table for the
{\bf S1} and {\bf S2} cases respectively), after imposing the
previously-mentioned cuts on the transverse momentum of the leading jet
$p_T(j_1) > 150$~GeV and on the parton-level missing
transverse energy $\slashed{E}_T > 150$~GeV. For both the {\bf S1} and {\bf S2}
scenarios, the NLO predictions are found to decrease with the dark matter mass,
paying the price of a phase-space suppression. The falling behaviour is found
steeper once the dark matter mass is greater than half the mediator mass, as it
has to be produced off-shell (\ie\ for $m_Y>2m_X$). Moreover, for given masses
and couplings, {\bf S1} cross sections (\ie\ in the case of mediator vector
couplings) are larger. This originates from the $p$-wave suppression of DM
production through an axial-vector mediator (\ie\ in the {\bf S2} scenario), as
already mentioned earlier in this section.

We then evaluate the cross section
value $\sigma_{95}$ that is excuded at the 95\% confidence level (third and
sixth columns of the table) by reinterpreting the results of the
ATLAS-CONF-2019-040 analysis. We observe that small dark matter masses are
excluded already with the full Run~2 dataset, cross sections as small as 100~fb
being excluded regardless of the DM mass. Moving on with a scenario in which the
couplings satisfy eqs.~\eqref{eq:scenario} and \eqref{eq:scenario3}, we
translate the bounds that we have obtained into bounds on a universal coupling.
The latter is found to be of at most in the 0.5--0.7 range once one lies in a
configuration below threshold ($2 m_X < m_Y$), and is mostly unconstrained for
larger DM mass values. As for the previous class of scenarios in which the DM
mass was fixed and the mediator mass was varying, 3000~fb$^{-1}$ will not improve
the limits much, as the analysis being dominated by the systematics. We indeed
expect an improvement on the bounds of at most 3-4\%.

\section{Conclusion}
\label{sec:conclusion}
In this paper we showcased new features of the \madanalysis\ package that improve the
recasting functionalities of the programme. These features focus on two aspects.

First, we have designed a way to include the uncertainties on the signal when
the code is used to reinterpret given LHC results in different theoretical
contexts. Theory errors on the total signal production cross section induced by
scale and PDF variations can be propagated through the reinterpretation
procedure. This results in an uncertainty band attached to the confidence level
at which a given signal is excluded. In addition, the user has the option to
provide information on the systematic uncertainties on the signal. With the
existence of new physics masses being pushed to higher and higher scales,
keeping track of error information on the signal becomes mandatory, especially
for what concerns the theoretical uncertainties, which can be significant for
beyond the Standard Model physics signals involving heavy particles.

Second, we have implemented a new option allowing the user to extrapolate the
constraining power of any specific analysis to a different luminosity, assuming
a naive scaling of the signal and background selection efficiencies. Several
options are available for the treatment of the background uncertainties,
depending on the information provided by the experimental collaborations for the
analyses under consideration. If information on the statistical and systematical
components of the uncertainties is available separately, signal region by signal
region, \madanalysis\ can use it to scale them up accordingly,
\ie~proportionally to the square root of the luminosity and linearly to the
luminosity for the statistical and systematical uncertainties respectively. In
contrast, if such
a detailed information is absent, the user is offered the choice to treat the
total error as being dominated either by statistics or by systematics, or in
his/her preferred fashion.

We have illustrated the usage of these new \madanalysis\ features  in the framework of three simplified models for new
physics.

First, we consider a signal that originates from the production of a
pair of gluinos that each decay into   two  jets and missing
transverse momentum. As an example, we  make predictions in
  the context of a simplified model inspired by the MSSM, in which only the
gluino and the lightest neutralino are light enough to be reachable at the LHC.
We have investigated the potential of   two ATLAS
searches for supersymmetry in 36~fb$^{-1}$  and
139~fb$^{-1}$  of LHC data. Those searches both rely on the effective mass variable and on the
presence of a large amount of missing transverse energy, and include
a large variety of signal regions featuring different jet multiplicity and
hadronic activity. We have reproduced to a
good approximation the ATLAS results at the nominal luminosity of the analysis
and compared our extrapolations at higher
luminosities with those obtained either through the more naive approach of the
{\sc Collider Reach} platform, or to publicly available ATLAS estimates for the
high-luminosity runs of the LHC. Fair agreement has been found.
 We have moreover studied the differences in the expected sensitivity
that arise when one considers, as a starting point, an analysis of 36 or
139~fb$^{-1}$ of Run~2 LHC data. Our predictions are found fairly compatible,
once theory errors are accounted for.

Second, we have focused on another MSSM-inspired simplified model in which the
SM field content is supplemented by one species of first generation squark and
one neutralino, all other supersymmetric states being decoupled. The spectrum
configuration is therefore such that the squark is heavier than the neutralino
and thus always decay into a light quark and a neutralino. This gives rise to a
multijet plus missing transverse energy signatures stemming from squark pair
production and decay. However, in contrast
with the gluino case, a smaller signal jet multiplicity is expected. We
have considered the same ATLAS supersymmetry search in 139~fb$^{-1}$ of LHC data
as above-mentioned, as it includes signal regions with a smaller jet
multiplicity so that some sensitivity to the considered simplified model is
expected. We have reinterpreted the results of the search and derived
up-to-date constraints on the model. We have then extrapolated our findings to
the high-luminosity LHC case.

Finally, we have considered an $s$-channel dark matter simplified model in which
one extends the Standard Model by a single dark matter candidate and one mediator that
connects the dark sector (made of the dark matter state) to the SM sector. We
have considered a fermionic dark matter state and a spin-1 mediator that couples
to a pair of SM quarks and a pair of dark matter particles. Typical dark matter
signals hence arise from the production of a pair of dark matter particles
(through an $s$-channel mediator exchange) in association with a hard visible
object. The most common case involves the production of a jet with a pair of
invisible dark matter particles, the signal being dubbed monojet in this case.
As the above ATLAS search is sensitive to such a signature (by virtue of the
properties of its low jet multiplicity signal regions), we reinterpret its
results to constrain the simplified model under consideration. We focus on and
compare two cases where the mediator couplings are of a vector and axial-vector
nature respectively. We then extract the current limits on the model, and
additionally project them at a higher-luminosity to get estimates for the LHC
sensitivity to the two studied $s$-channel dark matter setups.

 In all the models investigated , our results  emphasise the importance of embedding the uncertainties on
the signal. In one considered example, this could degrade the expected bounds
by about $10-20$\%, especially as a consequence of the large theory errors
originating from the poor PDF fit constraints at large Bjorken-$x$. Such a
regime is indeed relevant for new physics configurations still allowed by current
data and that involve the production of massive particles lying in the
multi-TeV mass range.

\section{Acknowledgements}
JYA acknowledges the hospitality of the LPTHE (Sorbonne Universit\'e) and IPPP
(Durham University) where parts of this work were completed.
MF acknowledges the NSERC for partial financial support under grant number
SAP105354.
BF has been supported by the LABEX ILP (ANR-11-IDEX-0004-02, ANR-10-LABX-63).

\bibliography{QCD-SUSY-Bounds}

\end{document}